\documentclass[fleqn,11pt]{wlscirep}
\usepackage[utf8]{inputenc}
\usepackage[T1]{fontenc}

\usepackage{graphicx}
\usepackage{bm}
\usepackage{siunitx}
\usepackage{physics}
\usepackage{color}
\usepackage[skip=5pt, indent = 15pt]{parskip}

\usepackage[font=footnotesize]{caption}
\usepackage[numbers, super, sort&compress]{natbib}
\usepackage{bibunits}
\defaultbibliographystyle{naturemag-doi}
\defaultbibliography{main}

\usepackage{hyperref}

\newcommand{\w}{{\mathcal D}}
\newcommand{\opt}{\mathrm{opt}}
\newcommand{\m}{\mathrm{min}}
\newcommand{\Wdis}{w_\mathrm{d}}

\newcommand{\naive}{\mathrm{naive}}
\newcommand{\gearshift}{\mathrm{gs}}
\newcommand{\kB}{k_\mathrm{B}}
\newcommand{\kBT}{\kB T}
\newcommand{\Cdf}{\Phi}
\newcommand{\cdf}{\phi}
\newcommand{\pideal}{p^*}

\def\thesection{\arabic{section}}
\def\thesubsection{\arabic{section}.\arabic{subsection}}
\makeatletter
\def\p@subsection{}
\def\p@subsubsection{}
\makeatother
\def\thesubsubsection{\arabic{section}.\arabic{subsection}.\arabic{subsubsection}}

\begin{document}
\title{Experimentally achieving minimal dissipation\\ via thermodynamically optimal transport}

\author[1]{Shingo Oikawa}
\author[1]{Yohei Nakayama}
\author[2, 3]{Sosuke Ito}
\author[4, 5]{Takahiro Sagawa}
\author[1,*]{Shoichi Toyabe}
\affil[1]{Department of Applied Physics, Graduate School of Engineering, Tohoku University, Sendai 980-8579, Japan}
\affil[2]{Department of Physics, Graduate School of Science, The University of Tokyo, Tokyo 113-0031, Japan}
\affil[3]{Universal Biology Institute, Graduate School of Science, The University of Tokyo, Tokyo 113-0031, Japan}
\affil[4]{Department of Applied Physics, Graduate School of Engineering, The University of Tokyo, Tokyo 113-8656, Japan}
\affil[5]{Quantum-Phase Electronics Center (QPEC), The University of Tokyo, Tokyo 113-8656, Japan}
\date{\today}

\begin{abstract} 
\end{abstract}

\maketitle

\begin{bibunit}

\noindent
\textsf{Optimal transport theory, originally developed in the 18th century for civil engineering\cite{Monge1781,Villani2009}, has since become a powerful optimization framework across disciplines, from generative AI\cite{Arjovsky2017,lipman2023flow} to cell biology\cite{Schiebinger2019}. In physics, it has recently been shown to set fundamental bounds on thermodynamic dissipation in finite-time processes\cite{Aurell2012, Nakazato2021}. This extends beyond the conventional second law, which guarantees zero dissipation only in the quasi-static limit and cannot characterize the inevitable dissipation in finite-time processes. Here, we experimentally realize thermodynamically optimal transport using optically trapped microparticles, achieving minimal dissipation within a finite time. As an application to information processing, we implement the optimal finite-time protocol for information erasure\cite{Aurell2012,Proesmans2020,Proesmans_2020-2}, confirming that the excess dissipation beyond the Landauer bound\cite{Landauer1961, Parrondo2015} is exactly determined by the Wasserstein distance --- a fundamental geometric quantity in optimal transport theory. Furthermore, our experiment achieves the bound governing the trade-off between speed, dissipation, and accuracy in information erasure. To enable precise control of microparticles, we develop scanning optical tweezers capable of generating arbitrary potential profiles. Our work establishes an experimental approach for optimizing stochastic thermodynamic processes. Since minimizing dissipation directly reduces energy consumption, these results provide guiding principles for designing high-speed, low-energy information processing.
}
\vspace{2mm}

\begin{figure*}[!htb]
    \centering
    \includegraphics{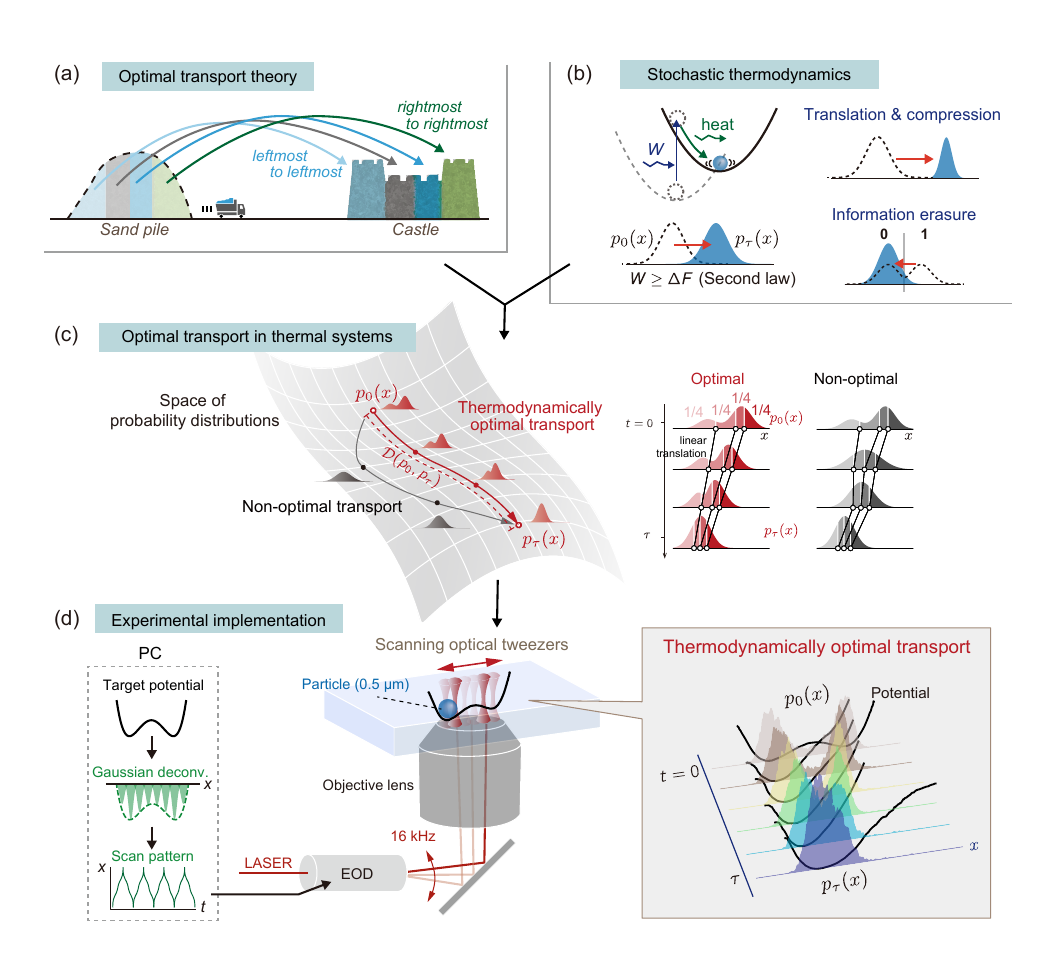}
    \caption{
    {\bf Optimal transport theory applied to thermal microparticles.}
    (a) Transport of a sand pile in one dimension.
    The protocol with minimizes the cost, defined based on the transport distance, is achieved when the positional order is maintained so that the sand grains at the leftmost location are transported to the leftmost location, and so on.
    Optimal transport is a transport protocol that minimizes the cost. When the cost is defined based on the transport distance, the optimal transport of moving a sand pile in one dimension is achieved when the positional order is maintained so that the sand grains at the leftmost location are transported to the leftmost location, and so on.
    (b) Stochastic thermodynamics describes the thermodynamics in thermal fluctuating systems \cite{Seifert2012, Ciliberto2017, Pigolotti-Peliti}. We think of transporting a probability distribution $p_0(x)$ at $t=0$ to $p_\tau(x)$ at $t=\tau$. Exemplified transport processes are shown on the right, which are the targets of this paper. 
    (c) Geometric space with the Wasserstein distance (left).
    With the optimal transport in one-dimensional systems, each segment in the distribution is linearly transported without changing the positional order (right).
    (d) We experimentally implement the optimal transport by using a microscopic particle with a diameter of \SI{0.5}{\micro m} trapped by a potential with a dynamically changing profile. We developed scanning optical tweezers that generate an arbitrary potential profile under the constraints determined by the device (SI Section~\ref{SI:Scanning optical tweezers}). Right: examples of transport. Distribution of the particle positions (color) and potentials reconstructed from the experiments (solid curves).}
    \label{fig:intro}
\end{figure*}

Consider the problem of transporting a pile of sand to another location (Fig.~\ref{fig:intro}a). In 1781, Gaspard Monge posed a deceptively simple yet fundamental question\cite{Monge1781}: Given the initial and final shapes of the sand piles and the cost of transporting each grain between any two positions, what is the most efficient way to minimize the total cost? This question in engineering laid the foundation for \textit{optimal transport theory}, which was later formalized in applied mathematics through the incorporation of probability theory. A central concept in this framework is the Wasserstein distance, which quantifies the difference between two probability distributions in terms of the minimal transportation cost required to transform one into the other\cite{Benamou2000, Villani2009}. In recent years, optimal transport theory has found applications across various disciplines, including thermodynamics. Notably, it has been shown to establish fundamental bounds on finite-time dissipation (Fig.~\ref{fig:intro}b, c), where the Wasserstein distance exactly characterizes the minimal dissipation in such processes\cite{ Aurell2012, Aurell2011, chen2019stochastic, Nakazato2021, ito2024geometric}.

Let us start with a simple scenario in which a microparticle is immersed in a thermal environment (heat bath) at temperature $T$. Due to thermal fluctuations, the particle undergoes stochastic motion, with its state described by a time-dependent probability distribution, denoted as $p_t(x)$. Here, $t$ represents time, and $x$ is the particle's position. Optimal transport theory provides a natural framework for optimizing the evolution of these probability distributions. Such stochastic thermodynamic systems form the foundation of modern thermodynamics --- often referred to as stochastic thermodynamics --- which applies not only to a microparticle but also to a wide range of experimental systems, including electric circuits and molecular motors\cite{Seifert2012, Ciliberto2017, Pigolotti-Peliti}.  

The second law of thermodynamics states that the thermodynamic work $W$ must always be greater than or equal to the nonequilibrium free-energy change $\Delta F$ \cite{Parrondo2015}, which implies that the dissipated work, defined as $\Wdis \equiv W - \Delta F$, satisfies $\Wdis \geq 0$. This thermodynamic dissipation, which is equivalent to the entropy production multiplied by $T$, vanishes only in the quasi-static limit requiring an infinitely long operation time. 
In finite-time processes, however, $\Wdis$ remains strictly positive due to unavoidable dissipation \cite{Shiraishi2016}. Optimal transport theory refines the second law by providing a universal bound on finite-time dissipation:  
\begin{align}\label{eq:Wdis:min}
    \Wdis \geq \gamma \frac{\w(p_0, p_\tau)^2}{\tau} \equiv \Wdis^\m,
\end{align}  
where the equality is achievable for any given duration $\tau$ \cite{Aurell2012, Nakazato2021}. Here, $\w(p_0, p_\tau) \geq 0$ represents the Wasserstein distance, which is determined solely by the initial and final probability distributions at times $t = 0$ and $\tau$, denoted as $p_0(x)$ and $p_\tau(x)$.
$\gamma$ is the particle's friction coefficient. A key feature of $\Wdis^\m$ is its inverse proportionality to $\tau$: the greater the speed $\propto 1/\tau$ is (i.e., the shorter the operation time $\tau$ is), the greater the additional dissipation is. 
In terms of geometry, the thermodynamically optimal transport that minimizes dissipation is realized by transport along a geodesic connecting $p_0$ and $p_\tau$ with a uniform velocity, as illustrated in Fig.~\ref{fig:intro}c (left) \cite{Villani2009,Benamou2000}.

A particularly important application of the second law of thermodynamics is in determining the fundamental energy cost of information processing \cite{Parrondo2015}. 
For example, the Landauer bound \cite{Landauer1961, Sagawa2009, Aurell2012, Lutz2015, Proesmans2020,Proesmans_2020-2} states that erasing one bit of information from a binary symmetric memory requires a minimum work of $W = \kB T \ln 2$, where $\kB$ is the Boltzmann constant.  
Since the Landauer bound is achieved only in the quasi-static limit, it is desirable to establish an achievable bound for finite time processes.  
This can be addressed using optimal transport theory: applying Eq.~\eqref{eq:Wdis:min} to information erasure yields  
\begin{align}\label{eq:Wdis:min:Landauer}
    W \geq \kB T \ln 2 + \gamma \frac{\w(p_0, p_\tau)^2}{\tau},
\end{align}  
where the additional term on the right-hand side vanishes in the limit $\tau \to \infty$.  
Finite-time bounds for information erasure and the corresponding optimal protocols have been theoretically obtained \cite{Aurell2012, Proesmans2020, Proesmans_2020-2}.  

Despite extensive theoretical studies on thermodynamic optimal transport, experimental validation has remained unaddressed due to the necessity of precisely controlling probability distributions.  
Achieving the fundamental bound in finite time requires implementing the optimal time evolution of the potential with high accuracy.  
For instance, in non-Gaussian transport processes such as information erasure, nonharmonic potentials must be precisely controlled. 
While the Landauer bound itself, in the limit $\tau \to \infty$, was experimentally demonstrated in 2012 using optically trapped microparticles \cite{Berut2012}, and various other experiments on the thermodynamics of information\cite{Jun2014, Gavrilov2016, RibezziCrivellari2019, Dago2021, Dago2023}, including implementations of Maxwell's demons have been conducted \cite{Toyabe2010, Koski2014},  
the thermodynamic optimization of information processing in finite time has been experimentally challenging.

In this study, we experimentally realize optimal transport by implementing the optimal protocols that minimize thermodynamic dissipation, providing the first proof-of-concept for thermodynamically optimal transport.
Our experimental platform consists of a Brownian microparticle confined in a dynamically controlled potential, serving as a prototypical thermodynamic system (Fig.~\ref{fig:intro}d; see also Extended Data Fig.~\ref{exfig:Optical tweezers}). To achieve the precise control required for optimizing distribution dynamics, we built a custom optical tweezer system capable of generating arbitrary potential profiles (within the constraints of the device) through precisely engineered laser scanning patterns (see Methods).  
This method is general and can be applied to a wide range of systems, including feedback control and simultaneous manipulation of multiple Brownian particles.

We first investigate a simple transport problem: the translation and compression of a Gaussian distribution.  
This scenario, due to its simplicity and experimental feasibility, provides a clear demonstration of optimal transport by allowing direct comparisons between optimal and non-optimal protocols.  
In particular, our experiment reveals the geometric structure of transport, showing that the optimal transport corresponds to uniform-speed motion along a geodesic in the space of probability distributions.  
Furthermore, we demonstrate that optimal transport theory provides a method to evaluate dissipated work based solely on the distribution dynamics, without requiring information of individual trajectories and potential profiles. This approach is applicable even to non-optimal protocols, and potentially to complex biological systems like molecular motors \cite{Toyabe2010PRL} and cells \cite{DiTerlizzi2024}.

Then, we perform the experiment on information erasure, which is the primary focus of this manuscript.  
To implement optimal information erasure, we dynamically vary the potential profile from an initial double-well configuration to a final single-well state, achieving the finite-time bound equivalent to Eq.~\eqref{eq:Wdis:min:Landauer}. 
Our experiment directly confirms that the finite-time correction to the conventional Landauer bound is given by the Wasserstein distance.

Another crucial aspect of information processing is accuracy.  
Typically, increasing speed increases dissipation (and thus energetic cost) and reduces accuracy \cite{Hopfield1974, Andrieux2008, Lan2012, Barato2015, dechant2022minimum,yoshimura2023housekeeping, 
Vu2023,ito2024geometric, Klinger2025}.
Such trade-offs between energy cost, speed (i.e., $1/\tau$), and accuracy are commonly observed in biological systems, including sensory adaptation\cite{Lan2012} and information replication\cite{Hopfield1974, Andrieux2008}.  
In our study, using the model experimental platform, we achieve the fundamental bound of this trade-off by implementing optimal information-erasure protocols with several different values of accuracy.
This demonstration reinforces the universality of such trade-off in thermodynamic information processing.

\section*{Results}

We begin by implementing a translation and compression protocol --- a simple yet highly controllable process --- to experimentally demonstrate and characterize optimal transport. Next, we realize the optimal transport for information erasure, marking the first experimental demonstration of finite-time thermodynamically optimal information processing. To accurately measure probability distributions and quantify physical quantities such as work, we perform extensive repetitions of each protocol, typically exceeding 12,000 repetitions, involving at least three different particles per condition (see Methods).

\subsection*{Optimal translation-compression transport in finite time}

Let $p_0$ and $p_\tau$ be Gaussian distributions with different means $\mu$ and standard deviations $d$.  
The Gaussian dynamics enable a detailed quantitative analysis of the transport process.  
To characterize optimal transport, we implement three distinct protocols: optimal, naive, and gearshift.

\begin{figure*}[!hbth]
    \centering
    \includegraphics{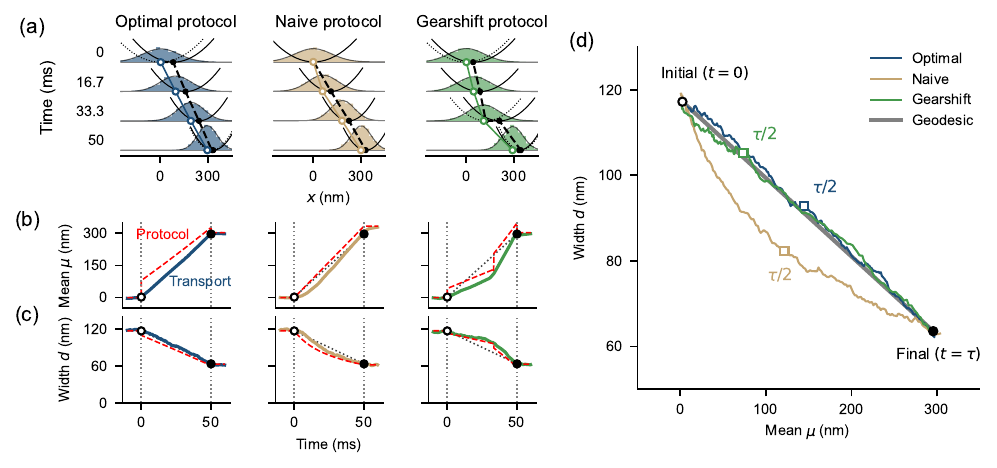}\vspace{-0.2cm}
    \includegraphics{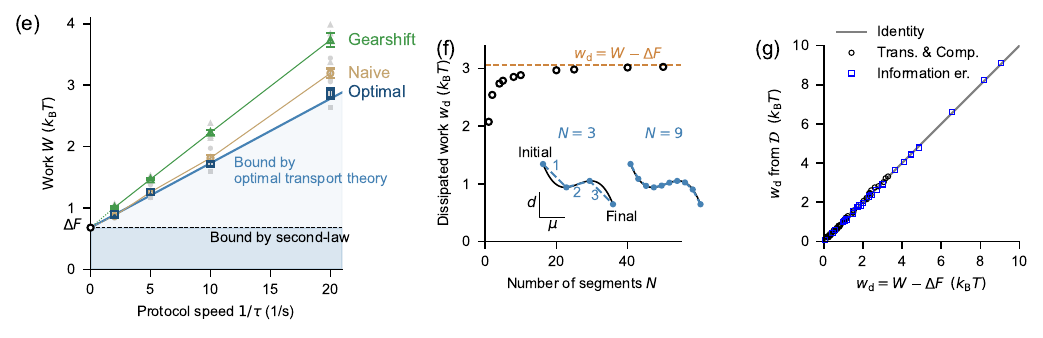}
    \caption{{\bf Optimal transport in finite time with translation and compression protocol. }
    (a -- c) Time evolution of probability distributions and potentials (a), mean $\mu$ (b), and width $d$ (c).
    The optimal protocol varies the potential profile so that $\mu_t$ and $d_t$ linearly vary.  
    The naive protocol linearly varies the position and stiffness of the potential. 
    The gearshift protocol combines two optimal protocols with different durations (fractions are 2/3 and 1/2) and speeds (ratio of 1 to 2).
    (a) Experimentally obtained distributions with Gaussian fittings and potentials for $\tau=\SI{50}{ms}$. Open and closed circles indicate the centers of distribution and potential, respectively.
    The dotted curves in optimal and gearshift protocols are the potentials before the jumps of the potential position. 
    (d) Trajectories in the ($\mu$, $d$) space, which implements the Wasserstein distance for Gaussian dynamics, for the same data in (a -- c). The optimal protocol is characterized by a uniform-speed transport on a geodesic (gray straight line) connecting the initial and final distributions.
    (e) The work $W$ vs the protocol speed $1/\tau$.
    $W$ was calculated based on Eq.~\eqref{eq:W, F}. 
 Gray closed symbol corresponds to an experimental run consisting of more than 3,000 repetitions for $\tau\le\SI{200}{ms}$ and 1,500 repetitions for $\tau=\SI{500}{ms}$ for a particle.
 We performed four runs with four independent particles under each condition to measure the mean values (colored open symbols).
    Error bars indicate the standard error of the mean (s.e.m., four samples).
    The black open circle indicates the mean values of $\Delta F$ calculated from the initial and final distributions.
    The blue solid line indicates the theoretical minimum evaluated using the mean $\Delta F$ (0.680 $\pm$ 0.007, mean $\pm$ s.e.m. of all data of all protocols, 48 samples) as the intercept and the mean of $\tau\Wdis^\m$ with $\Wdis^\m$ calculated by Eq.~\eqref{eq:Wdis:min} as the slope.
    Some runs show $W$ values lower than this average theoretical minimum (also in Figs.~\ref{fig:landauer}d and \ref{fig:tradeoff}), since the minimum $\Delta F+\Wdis^\m$ differs from particle to particle even in the same condition due to the particle-dependent variation in $\gamma$ (Extended Data Fig.~\ref{exfig:power spectrum}). We confirmed that each run satisfies the bound except for a few outliers due to statistical errors (Extended Data Fig.~\ref{exfig:Wdis:each}).
    The colored thin solid lines connect experimental data of naive and gearshift protocols, which are extrapolated to the circle by dotted lines.
    (f) Evaluation of $\Wdis$ from distributions without knowing individual trajectories (Eq.~\eqref{eq:Wdis:W2}).  A typical example of gearshift protocol is shown. Inset: schematic of the segmentation.
    (g) Comparison of evaluation of $\Wdis$ from recovered potentials (Eq.~\eqref{eq:W, F}) and from distributions (Eq.~\eqref{eq:Wdis:W2}). 
        }
    \label{fig:translation}
\end{figure*}

We first constructed the {\it optimal} protocol for given $p_0$ and $p_\tau$ (Fig.~\ref{fig:translation}a-d). 
If $p_0$ and $p_\tau$ are both Gaussian, the intermediate distributions under the optimal transport protocol are always Gaussian, with linearly varying $\mu$ and $d$ ~\cite{Nakazato2021} for a distance of $\mu_\tau-\mu_0=\SI{300}{nm}$ and a compression ratio of $d_0/d_\tau = 2$.
$p_0$ and $p_\tau$ are chosen to be the same as the following naive protocol.
The dynamics of potential $V_t(x)$ realizing the transport are obtained by numerically solving the Fokker-Planck equation (see SI Section~\ref{SI:Computation of potential}), which is directly implemented in our experiment.
$V_t(x)$ is always harmonic and has a discrete forward jump of the parameters at $t=0$ and a backward jump at $t=\tau$ (Fig.~\ref{fig:translation}a-c).
The first jump compensates for the delay due to viscous relaxation, and the last jump quenches the dynamics to the final target distribution.

Transport can be geometrically characterized in the distribution space. 
We observed that the designed optimal protocol realizes the linear translation of the distribution in both $\mu$ and $d$ (Fig.~\ref{fig:translation}a -- c).
Accordingly, we obtained a linear uniform-velocity trajectory in the $(\mu, d)$ space (Fig.~\ref{fig:translation}d), where the Euclidean distance is equal to the Wasserstein distance for Gaussian distributions~\cite{Villani2009}. 
The uniform-velocity transport on a geodesic in the distribution space indicates the optimal transport~\cite{Benamou2000,Nakazato2021}.

The {\it naive} protocol was implemented as a reference, where the position and stiffness of a harmonic potential are linearly varied.
The particle followed the potential with a time delay owing to viscous relaxation.
Therefore, the final position and width of the distribution at $t=\tau$ do not reach the equilibrium values for the potential at $t=\tau$.
The trajectory in the $(\mu, d)$ space significantly deviated from that of the optimal protocol (Fig.~\ref{fig:translation}d).

As a further reference, we also attempted a {\it gearshift} protocol, which connects two optimal protocols with different durations and speeds.
This protocol realized a transport on the geodesic similarly to the optimal protocol but with a non-uniform speed (Fig.~\ref{fig:translation}d).
In this sense, the protocol is not optimal as a whole.

\subsubsection*{Work}

The work $W$ and free energy change $\Delta F$ for transport are evaluated by using the potential $V_t$, recovered from the experimental trajectories (see Methods), and the distribution $p_t$ (Fig.~\ref{fig:translation}e).
The optimal protocol achieves the theoretical minimum for finite time processes given by Eq.~\eqref{eq:Wdis:min} within error bars.
Accordingly, the energy-speed tradeoff $\Wdis\propto 1/\tau$ was observed.
$\Delta F$ was 0.680 $\pm$ 0.007 $\kBT$ (mean $\pm$ s.e.m. of all data, 48 samples).
This corresponds to the compression ratio of $\exp(\Delta F/\kBT)=1.97$, which is close to the designed value of 2.
On the other hand, the naive protocol has larger $\Wdis$ and has a slightly nonlinear dependence on $1/\tau$; this implies that the transport is in the nonlinear-response regime.
For a systematic comparison, we also constructed intermediate protocols by linearly interpolating optimal and naive protocols (Extended Data Fig.~\ref{exfig:nonoptimality}).

The $1/\tau$ dependence is also observed with the gearshift protocol.
This is because the trajectories in the $(\mu, d)$ space are similar for different $\tau$.
However, $W$ did not reach the theoretical minimum, indicating that $\Wdis\propto 1/\tau$ alone does not necessarily indicate optimal transport.

\subsubsection*{Evaluation of dissipated work without knowing the potential}

The work corresponds to the energy change resulting from the change in the shape of $V_t(x)$ 
 ~\cite{Pigolotti-Peliti}.
Therefore, it is straightforward to use $V_t(x)$ to calculate dissipated work $\Wdis$ based on Eq.~\eqref{eq:W, F} in Methods as practiced above.
However, recovering $V_t(x)$ requires a large set of trajectories and thus is not always feasible in experiments, especially if treating complex systems such as biological systems.
In contrast, the optimal transport theory allows the calculation of $\Wdis$ only from the distribution dynamics $p_t(x)$ during the process (in the absence of nonconservative force), without using information about the potential profile $V_t (x)$ ~\cite{Nakazato2021}.  
This method does not require individual trajectories, and furthermore, is applicable regardless of whether the process is optimal or not (see Methods).

Consider dividing the time interval $[0, \tau]$ into $N$ short segments.
In the $(\mu, d)$ space, the transport in each segment is approximated by linear transport with uniform speed, that is, the optimal transport if the segment is sufficiently short (Fig.~\ref{fig:translation}f).
Thus, $\Wdis$ of the whole process is estimated as the sum of $\Wdis^\m$ in each segment.
This method assumes the absence of the non-conservative force, which is always the case in one dimension~\cite{Nakazato2021}.

We found that $\Wdis$ computed by this method converges to the value computed using $V_t(x)$ at large $N$, validating the methodology (Fig.~\ref{fig:translation}f, g).
The number of segments $N$ needed for convergence is determined by the curvature and uniformity of the velocity of the whole transport trajectory in the $(\mu, d)$ space.

\subsection*{Optimal information erasure in finite time}

\begin{figure*}[bth]
    \centering
    \includegraphics{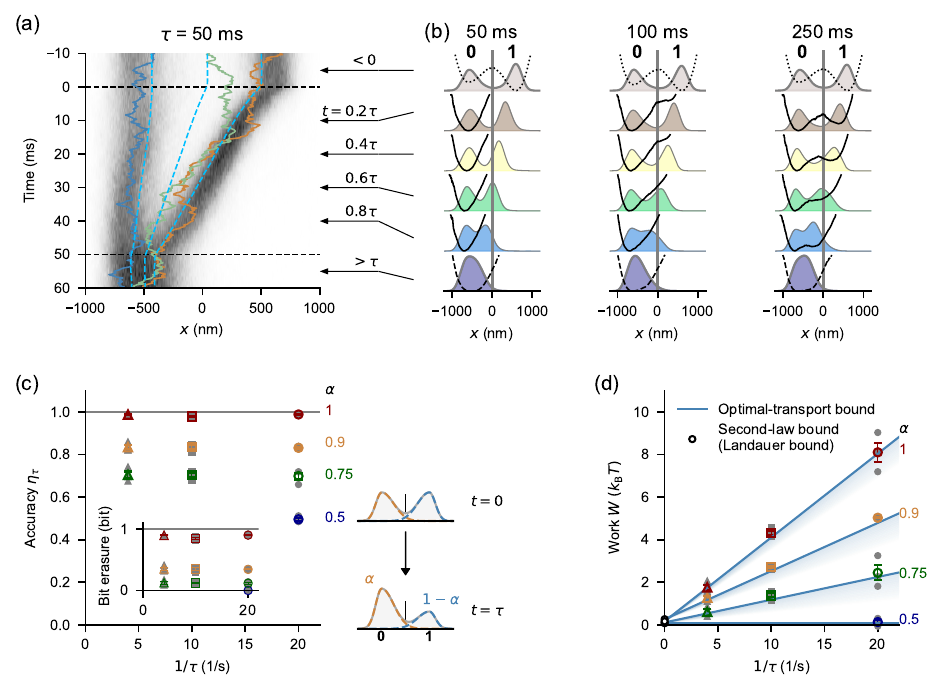}
    \caption{{\bf Optimal information erasure in finite time. }
    (a) Kymograph of the probability distributions constructed from 5,585 repetitions of information erasure with exemplified trajectories (solid).
    The cyan dashed curves indicate the tertile and mean of the distribution.
    (b) The distribution $p_t(x)$ and the recovered potential $V_t(x)$ under the optimal protocol.
    The optimal potential dynamics changed instantaneously at $t=0$ and $t=\tau$, similarly to the translation-compression setup.
    Each distribution is calculated from 31 successive video frames and spatially smoothened by being convolved with a Gaussian-shape window.
    (c) Accuracy of information erasure $\eta_\tau$ evaluated as the fraction of \textbf{0} at $t=\tau$.
    The inset is the bit erasure calculated as $\Delta H\times \log_2 e$ plotted against $1/\tau$.
    $\alpha$ is a parameter to control the accuracy and is the height ratio of the two peaks in the final target distribution.
    With $\alpha=0.5$, the potential is unchanged during the transport.
    (d) Work. Solid lines correspond to the theoretical minimum for work $\Delta F+\Wdis^\m$, where we use the mean $\tau\Wdis^\m$ for each $\alpha$ as the slope and the mean $\Delta F$ for each $\alpha$ as the intercept. 
    Number of samples (particles) is three for each point in (c) and (d).
     Gray closed symbols correspond to each run of more than 5,000 repetitions.
    Colored open symbols are the mean of each condition.
    Error bars indicate s.e.m. (three samples for each).
    }
    \label{fig:landauer}
\end{figure*}

We now turn to the experiment on optimizing information erasure in finite time.  
Specifically, we consider a situation where one bit of information is encoded in a symmetric double-peak distribution, with logical state \textbf{0} assigned to $x < 0$ and \textbf{1} assigned to $x \geq 0$ (Fig.~\ref{fig:landauer}).  
The information erasure process transforms the double-peak distribution into a single-peak distribution corresponding to a fixed logical state.  
Without loss of generality, we focus on resetting to logical state \textbf{0}, as the symmetric double-peak ensures the symmetry between \textbf{0} and \textbf{1}.

We experimentally implemented the optimal information erasure protocol that was obtained numerically (Fig.~\ref{fig:landauer}).
The kymograph clarifies the distribution dynamics (Fig.~\ref{fig:landauer}a).
The protocol translates the fraction of the distribution in the state \textbf{1} to the state \textbf{0}. 
The fraction in \textbf{0} is slightly compressed leftward to save space for the incoming fraction from \textbf{1}.
As a result, we observed a linear variation of the tertiles and mean of the distribution (dashed curves in Fig.~\ref{fig:landauer}a).
This is the characteristic of the optimal transport as shown in Fig.~\ref{fig:intro}c (right).
The optimal transport dynamics are similar for different $\tau$ when time is scaled by $\tau$ (Fig.~\ref{fig:landauer}b).
This is also the characteristic of optimal transport and is realized by different potential dynamics depending on $\tau$.

The accuracy of the information erasure is measured by the fraction of the state \textbf{0} at $t=\tau$, denoted as $\eta_\tau = \int^{0}_{-\infty} p_{\tau} (x)\dd x$.
An almost perfect erasure with $\eta_\tau=0.984\pm 0.005$ (mean $\pm$ standard deviation (s.d.)) was achieved even within finite time (Fig.~\ref{fig:landauer}c, $\alpha =1$).
Because the target final distribution has a tail extending beyond $x=0$, perfect transport is not always expected.
The corresponding bit erasure was $0.88 \pm 0.03$ bit (mean $\pm$ s.d., Fig.~\ref{fig:landauer}c, inset), which was quantified as $\Delta H\times \log_2 e$.
Here, $H(\eta) = -\eta\ln\eta -(1-\eta)\ln (1-\eta)$ is the Shanon information content defined in the natural logarithm, and $\Delta H=H(\eta_0) - H(\eta_\tau)$.
$\eta_0 =\int^{0}_{-\infty} p_{0} (x)\dd x$ was $0.495\pm 0.009$ (mean $\pm$ s.d.).

\subsubsection*{Work}

We measured the work $W$ during the information erasure process (Fig.~\ref{fig:landauer}d, $\alpha = 1$).
$W$ reached the finite-speed theoretical minimum given by $\Delta F+\Wdis^\m$ within error bars, validating the realization of optimal finite-speed information erasure.
$\Wdis^\m$ is given by Eq.~\eqref{eq:Wdis:min}.
The free energy difference $\Delta F$ corresponds to the Landauer bound, which can be reached in the quasi-static limit ($1/\tau\to 0$).
$\Delta F$ consists of the free energy change due to the bit erasure, $\kBT\Delta H$ ~\cite{Landauer1961, Berut2012}, and the rearrangement of the particle distribution inside the \textbf{0} and \textbf{1} states \cite{Sagawa2014}.

The values of $\Wdis$ evaluated solely from the distributions coincided with those from the recovered potential (Fig.~\ref{fig:translation}g), again validating the effectiveness of the distribution-based evaluation of $\Wdis$ with this non-harmonic setup.

\subsubsection*{Energy-speed-accuracy trade-off}

It is generally expected that more accurate control requires more work, and faster control reduces accuracy, implying the trade-off between energy cost $\Wdis$, speed $1/\tau$, and accuracy $\eta_\tau$.
To control the accuracy, we left a fraction of the distribution at $t=\tau$ so that the final distributions have double peaks; the height ratios of the two peaks are $\alpha$ to $1-\alpha$ ($0.5\le\alpha\le 1$, see the right panel of Fig.~\ref{fig:landauer}c and SI Section \ref{SI:Information erasure}).
The distributions are designed so that they are approximately local equilibrium distributions in each well of a double-well potential.
The accuracy $\eta_\tau$ increases with $\alpha$. However, $\alpha$ does not solely determine $\eta_\tau$, since the peaks have tails extending beyond $x=0$ as mentioned.
We observed that the work become smaller with smaller $\alpha$ as well as smaller $1/\tau$ (Fig.~\ref{fig:landauer}d), which implies the trade-off between energy cost, speed, and also accuracy.
That is, a faster and more accurate process requires more work.

\begin{figure*}[bth]
    \centering
    \includegraphics{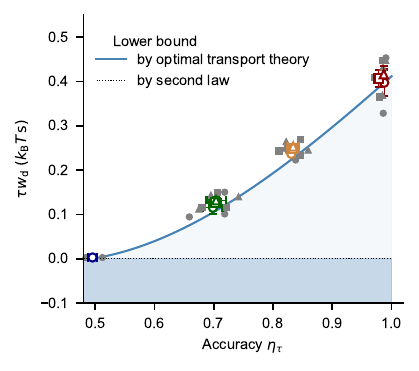}
    \caption{{\bf Trade-off between energy cost, speed, and accuracy.}
    The dissipated work $\Wdis$ was multiplied by $\tau$ to illustrate the trade-off, since $\Wdis$ scales with $1/\tau$ for optimal transport (Eq.~\eqref{eq:Wdis:min}).
    The solid curve indicates the bound by optimal transport theory (Eq.~\eqref{eq:Wdis:min}).
    The symbols are the experimental data.
    The bound curve was constructed by interpolating the mean of $\tau\Wdis^\m$ for the data with the same $\alpha$ (indicated by the same colors) by a cubic spline curve.
    The dotted line corresponds to the bound by the second law.
    See also Extended Data Fig.~\ref{exfig:tradeoff}.
     The colors and symbols are the same as those in Fig.~\ref{fig:landauer}d.
      Gray closed symbols correspond to each experimental run of more than 5,000 repetitions.
    Colored open symbols are the mean values in independent runs in each condition (three samples for each).
The error bars indicate s.e.m..
    }
    \label{fig:tradeoff}
\end{figure*}

Figure \ref{fig:tradeoff} shows our experimental data in a way to clarify that they achieve the bound of the energy-speed-accurary trade-off.
The values of $\tau\Wdis$ for different $\tau$ collapsed into a single curve, which corresponds to the finite-time minimum $\gamma\w(p_0,p_\tau)$ predicted by the optimal transport theory (solid line, Eq.~\eqref{eq:Wdis:min}).
The fact that $\gamma\w(p_0,p_\tau)$ has a finite value independent of $\tau$ indicates that $\tau\Wdis$ does not reach zero even in the quasi-static limit $\tau\to\infty$.
The ordinary second law only claims the positivity of $\Wdis$ (dotted line).
Since $\eta_\tau$ depends on $\w(p_0,p_\tau)$, the results demonstrate the trade-off between $1/\tau$, $\Wdis$, and $\eta_\tau$. 

\section*{Discussion}

In this study, we experimentally demonstrated thermodynamically optimal transport by implementing protocols that minimize dissipated work $\Wdis$. We built a custom optical tweezer system capable of generating arbitrary potential profiles to optimize the distribution dynamics of Brownian microparticles in a thermal environment. We first demonstrated a simple transport problem of translating and compressing a Gaussian distribution, revealing the geometric structure of transport (Fig.~\ref{fig:translation}). We then experimentally applied the optimal transport protocol to information erasure, achieving the finite-time Landauer bound, equivalent to Eq.~\eqref{eq:Wdis:min:Landauer} (Fig.~\ref{fig:landauer}). Our experiment achieved the trade-off bound between energy cost, speed, and accuracy (Fig.~\ref{fig:tradeoff}).

An approach that appears similar but is fundamentally distinct from optimal transport is \textit{optimal control} \cite{Schmiedl2007, Blaber2023, Loos2024}.  
While optimal transport directly optimizes the evolution of probability distributions, optimal control focuses on optimization through a certain set of potential parameters and thus does not necessarily produce the desired final distribution in finite time.  
This distinction becomes particularly relevant in the context of information erasure in finite time, where the erased information content is solely determined by the final compressed distribution but not by the specific profile of the final potential.  
In general, optimal control protocols differ from optimal transport protocols, as they are derived based on different optimization strategies (see also Extended Data Fig.~\ref{exfig:OptimalControl} and SI Section \ref{SI:optimal control}).

We note that there is yet another approach to finite-time thermodynamic trade-offs, called thermodynamic uncertainty relations (TURs) \cite{Barato2015, Horowitz2019}, which have been tested and used to estimate dissipation from experimental data \cite{Song2021, Marsland2019}.  
However, the bounds provided by TURs are often unachievable through experiments.  
In contrast, the framework based on optimal transport theory provides an achievable bound (Eq.~\eqref{eq:Wdis:min}) along with its optimal protocol, as experimentally demonstrated in this study.

Meanwhile, modern computers generate vast amounts of dissipation \cite{Ball2012, Markov2014}.  
In the long term, their energetic efficiency will be fundamentally constrained by thermodynamic laws, such as the Landauer bound \cite{Landauer1961, Parrondo2015, Pigolotti-Peliti} and its finite-time refinement (Eq.~\eqref{eq:Wdis:min:Landauer}).
Our experiment highlights the crucial role of optimizing temporal dynamics in approaching such fundamental bounds.  
Given that CMOS technology underpins modern computing and operates far from the quasi-static limit, a fundamental challenge is whether its architecture can achieve such thermodynamic bounds \cite{Freitas2021, Wolpert2024}.  
While our study serves as a proof-of-concept, it is expected to provide guiding principles for the design of more energy-efficient computing devices.

\section*{Acknowledgements}

We thank Takayuki Ariga and Kenji Nishizawa for their technical assistance.
This work was supported by JST ERATO Grant Numbers JPMJER2204 and JPMJER2302, and JSPS KAKENHI Grant Numbers 22H01141, 23H01136, 23H00467, and 24H00834.

\section*{Author Contributions}

SO, YN, SI, TS, and ST designed the research and wrote the paper. SO performed experiments. SO, YN, and ST developed the experimental systems, contributed analytic tools, and analyzed data.


\end{bibunit}

\newpage

\section*{Methods}

\renewcommand{\theequation}{M\arabic{equation}}
\setcounter{equation}{0}

\setcounter{section}{0}  
\setcounter{subsection}{0} 
\setcounter{subsubsection}{0} 

\begin{bibunit}
\renewcommand{\bibnumfmt}[1]{M#1.}
\renewcommand{\citenumfont}[1]{M#1}

\subsection*{Experimental setup}

An infrared laser with a wavelength of \SI{1064}{nm} (Spectra-Physics (MKS Instruments), MA) was focused through a 100$\times$ objective lens (NA1.40, Evident, Japan), specialized for a near-infrared laser, equipped to an inverted microscope (Evident) to create an optical trap (Extended Data Fig.~\ref{exfig:Optical tweezers}).
The laser power was adjusted by an attenuator (ThorLabs, NJ).
The typical laser power at the sample was \SI{3}{mW}, which was measured by an optical power meter (ThorLabs).

We trapped a silica particle with a diameter of \SI{500}{nm} (Micromod, Germany) diluted by distilled water in an observation chamber with a height of \SI{20}{\micro m}.
The trap position was approximately \SI{6}{\micro m} from the bottom glass surface.
The chamber was made by sticking two pieces of coverslips (Matsunami, Japan) together with double-sided tape (Teraoka, Japan).
The inlet and outlet of the channel were sealed with nail polish (DAISO, Japan) to prevent evaporation.
The particle images were taken by a high-speed camera (Basler, Germany) at \SI{4}{kHz} with an exposure time of \SI{60}{\micro s} under LED illumination (ThorLabs).
The room temperature was $24\pm 1$ \si{\degreeCelsius}.
The experiments were controlled by LabVIEW software (NI, TX).

The laser focal point was scanned by an electro-optical deflector (Conoptics, CT) at \SI{16}{kHz} to create a trapping potential.
The translation speed was controlled so that the mean light intensity at each position is proportional to the designed value of the potential at each position.
We deconvolved target potential profiles by Gaussian intensity profile that approximates the laser spot to obtain the scan pattern under constraints that the total power is fixed, the spatial scanning range is limited, and the mean time duration residing at each position is positive (Extended Data Fig.~\ref{exfig:Scan}, see SI Section~\ref{SI:Scanning optical tweezers} for details).

We repeated more than 3,000 repetitions for $\tau\le\SI{200}{ms}$ and 1,500 repetitions for $\tau=\SI{500}{ms}$ in each run of the translation-compression protocols for a single particle and 5,000 repetitions in each run of the information-erasure protocols for a single particle.
We conducted at least three runs with three different particles under each condition (actual numbers are specified in the figure captions). 
Each cycle of the repetitions consists of the following steps (Extended Data Figs.~\ref{exfig:ScanPattern:translation} and \ref{exfig:ScanPattern:landauer}); initial manipulation, pre-transport equilibration, transport, and post-transport equilibration.
The initial manipulation step is only used in the information erasure, which is intended for fast relaxation to the equilibrium of the initial state between \textbf{0} and \textbf{1}.

The particle position was evaluated as the centroid of the particle image, 
$(X, Y) = [\sum_{i, j}(s_{i, j} - s_\mathrm{th})x(x_i, y_i)] / \sum_{i, j}(s_{i, j} - s_\mathrm{th})$, where $s_{i, j}$ is the pixel intensity at position $(i, j)$.
The threshold intensity $s_\mathrm{th}$ is determined as the top 20\% of the intensity distribution of the whole image.
This fraction-based thresholding is expected to reduce the noise due to the temporal illumination variation.
The sum was taken for the pixels in the largest cluster of the pixels with intensities larger than $s_\mathrm{th}$, which was further processed by erosion and dilation, to reduce the effect of noise.
The precision evaluated as the s.d. of the centroid of a particle fixed on a glass surface was \SI{6.2}{nm}.
This value is a composite value including other effects such as the oscillation of the camera, microscope body, and microscope stage.

\subsection*{Wasserstein distance and transport protocols}

Think of transporting a one-dimensional distribution $p(x)$ to $q(x)$.
The transport map is expressed as ${\cal A}_{p\to q}(x)$ such that $q(x)=\int\dd x'p(x')\delta(x-{\cal A}_{p\to q}(x'))$.
For given two one-dimensional distributions $p(x)$ and $q(x)$, 2-Wasserstein distance $\w(p,q)$ is defined as
\begin{align}\label{eq:Wasserstein}
    \w(p,q)^2=\min_{{\cal A}_{p\to q}}\int\dd x ||x-{\cal A}_{p\to q}(x)||^2p(x),
\end{align}
subject to the Jacobian equation $\left|\frac{\partial {\cal A}_{p\to q}(x)}{\partial x}  \right|q({\cal A}_{p\to q}(x))= p(x)$ ~\cite{Villani2009}.
Here, $||x-y||$ denotes an Euclidean distance and is $|x-y|$ in one-dimensional systems.
We used a Python library ``POT: Python Optimal Transport'' ~\cite{flamary2021pot} for calculating the Wasserstein distance.

For an overdamped Langevin dynamics, the minimum transport cost is given by Eq.~\eqref{eq:Wdis:min}.
The optimal transport that achieves this minimum is numerically obtained \cite{Benamou2000}.
In one-dimensional Euclidean space, the optimal transport is a linear transport without changing the positional order, such as leftmost to leftmost and center to center~\cite{Villani2009} (Fig.~\ref{fig:intro}c, right).
The potential dynamics that realize this optimal transport are obtained by numerically solving the Fokker-Planck equation.
See SI Section~\ref{SI:Optimal protocol} for details, including the naive, gearshift, and intermediate protocols.

\subsection*{Evaluation of potential and work}

The potential profile was recovered based on a drift velocity.
By discretizing the Langevin equation $\gamma\dot x=f(x)+\sqrt{2\gamma\kBT}\xi$, we obtain $\gamma(x_{i+1}-x_i)=f(x_i)\Delta t+\sqrt{2\gamma\kBT}(B(t_{i+1}) - B(t_i))$.
Here, $f(x) = -\frac{\partial V}{\partial x}$ is the potential force, and $\xi$ is the white Gaussian noise with zero mean and unit variance. $B(t)$ is a Wiener process. We obtain $f(x_i)\Delta t/\gamma$ by splitting $x_i$ into spatial bins and calculating the average of $x_{i+1}-x_i$ in each bin, since the mean of $B(t_{i+1}) - B(t_i)$ is zero.
$\gamma$ was estimated as described below.
Then, $V(x_i)$ is recovered by integrating $f(x_i)$ and then smoothed by a window averaging.

This method is applicable to trajectories that are not settled in equilibrium.
We applied the method to the trajectories during transport, where the potential profile varies in time.
For each video frame, we use multiple consecutive frames around that frame of all the repetitions in each run for better statistics to obtain the potential profile.
We used 21 frames for $\tau = \SI{50}{ms}$ and $\SI{100}{ms}$ and 41 frames for $\tau=\SI{250}{ms}$.
Slower dynamics with longer $\tau$ allow us to use more frames.
Extended Data Figure~\ref{exfig:potential} shows examples of the recovered potentials, which are quantitatively similar to the target potentials.
The potential dynamics realized by the optical tweezers are constrained by the diffraction limit.
The fact that the minimum dissipated work can still be obtained suggests that the rough transport design determines the dissipated work, and the specific details do not significantly affect it.
This tolerance implies the effectiveness of the optimal transport theory in broad practical systems.

The dissipated work $\Wdis=W-\Delta F=W-F_\tau+F_0$ is calculated using \cite{Seifert2012, Pigolotti-Peliti}
\begin{align}\label{eq:W, F}
\begin{split}
W&=\int_0^\tau \dd t\int^\infty_{-\infty} \dd x \pdv{V_t(x)}{t}p_t(x),\qquad 
F_t = \int^\infty_{-\infty}\dd xV_t(x)p_t(x) - TS_t,\\
    S_t&=-\kB\int^\infty_{-\infty} \dd x p_t(x)\ln p_t(x).
\end{split}
\end{align}
We calculated these values based on experimental trajectories as follows.
Let $x_k$ and $V_k(x)$ be the particle position and potential in the $k$-th frame, respectively.
The transport step corresponds to $1\le k\le L$. $k=0$ and $k=L+1$ correspond to the last frame of the pre-transport step and the first frame of the post-transport step, respectively.
$W$ was calculated as 
\begin{align}
W=\left\langle\sum^{L}_{k=0} \left[V_{k+1}(x_{k+1})-V_{k}(x_{k+1})\right]\right\rangle.
\end{align}
$V_{0}$ and $V_{L+1}$ are the potentials before and after the transport process, respectively.
$\langle\cdot\rangle$ denotes the average between different repetitions.
We obtained $\Wdis=W-\Delta F$ by calculating
$\Delta F=\Delta V-T\Delta S$, $\Delta V=\langle V_{L+1}(x_{L+1})-V_0(x_0)\rangle$, and $\Delta S=\kB\sum_jp_{L+1,j}\ln p_{L+1,j}-\kB\sum_jp_{0,j}\ln p_{0,j}$.
Here, $j$ specifies the spatial bin, and $p_{k,j}$ is the probability of being in the $j$-th bin at $k$-th frame.
For the evaluation of $p_{k, j}$, multiple frames (11 frames) around a target frame were used for better statistics.

\subsection*{Evaluation of the dissipated work from distribution dynamics}

Consider dividing $p_t(x)$ into $N$ short transport segments with time duration $[t_i, t_{i+1}]$ ($i=1, 2, \ldots, N)$ (Fig.~\ref{fig:translation}f, inset).
The dissipated work during $i$-th segment, denoted as $w_i$, is bound as
\begin{align}\label{eq:sigma:min:i}
w_i\ge \gamma\frac{\w(p_{t_i}, p_{t_{i+1}})^2}{t_{i+1}-t_i}\equiv w^\m_i.
\end{align}
By taking the summation over $i$, we obtain $\Wdis\ge \sum_{i=1}^{N}w^\m_i$.
In the limit of $t_{i+1}-t_i\to 0$, we expect that $w_i$ converges to $w^\m_i$ since $p_{t_i}(x)\simeq p_{t_{i+1}}(x)$ if we consider the one-dimensional Euclidean space where the nonconservative force does not exist~\cite{Nakazato2021}.
Hence,
\begin{align}\label{eq:Wdis:W2}
\Wdis=\lim_{N\to\infty}\sum_{i=1}^{N}w^\m_i.
\end{align}

Equation \eqref{eq:Wdis:W2} enables us to evaluate $\Wdis$ only from the distribution dynamics without knowing the potential profiles or individual trajectories.
Figure~\ref{fig:translation}f, g demonstrates the validity of the method.
Extended Data Figure~\ref{exfig:Methods:Wdis:cycleN} shows the dependence of $\Wdis$ evaluated from the distributions on the number of repetitions.
The value of $\Wdis$ converges from above to a specific value.
More than 3,000 repetitions are necessary for sufficient convergence in the present setup in the information erasure.
We measured more than 5,000 repetitions, which is sufficient for the convergence.

\subsection*{Evaluation of friction coefficient}

The friction coefficient $\gamma$ of each particle was measured before the transport experiments.
The power spectrum of the particle position $x(t)$ obeys a Lorentzian spectrum in a harmonic potential (Extended Data Fig.~\ref{exfig:power spectrum}):
\begin{align}\label{eq:Lorentzian}
    C(f) = \int^\infty_{-\infty}\langle x(t)x(0)\rangle e^{2\pi if t}dt= \frac{\kBT}{2\pi^2\gamma}\cdot\frac 1{f^2+f_0^2},
\end{align}
where $f$ is a frequency, $f_0=k/(2\pi\gamma)$ is a corner frequency, and $k$ is the trap stiffness.
$\gamma$ is obtained by least-square fitting of $f C(f)$ with the fitting parameters $f_0$ and $\gamma$.
The multiplication by $f$ biases the fitting weight to the frequency region around $f_0$, which is intended to improve the fitting accuracy.
The value of $\gamma$ was $1.01\pm 0.03\,\, \kBT\,\si{s/\micro m^2}$ (mean$\pm$s.d.).
This value is similar to that estimated by Stokes law, $\gamma=6\pi\eta a = 1.05\,\, \kBT\,\si{s/\micro m^2}$, where $\eta=\SI{0.911}{mPa\, s}$ at \SI{24}{\degreeCelsius} (average room temperature) is the viscosity of water, and $a=\SI{0.25}{\micro m}$ is the particle radius.
Since the precise value of $a$ is also not known, the estimation by the Stokes law was used only as a reference.

\clearpage
\renewcommand{\figurename}{Extended Data Figure}
\setcounter{figure}{0}  

\noindent\textsf{\textbf{\Large Extended Data Figures}}

\begin{figure}[htbp]
    \centering
    \includegraphics{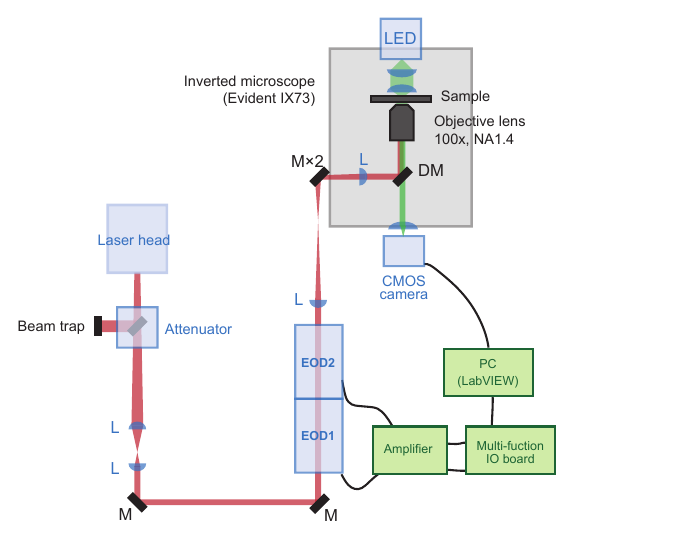}
    \caption{{\bf Optical tweezers system.} Laser: SpectraPhysics V-106C-4000 (\SI{4}{W} at maximum, \SI{1064}{nm}). Electric optical deflector (EOD) and amplifier: ConOptics 412-2Axis system. Attenuator: ThorLabs VA5-1064/M. Beam trap: ThorLabs BT610/M. CMOS camera: Basler ace acA1300-200.
    Microscope: Evident IX73. Objective lens: Evident UPlanSApo (100$\times$, NA1.40) customized for infrared.
    Multi-function IO board: NI PCIe-6374. 
    L and M denote lens and mirror, respectively.
    M$\times$2 means that there are two overlapping mirrors that reflect light in the direction perpendicular to the paper and in the direction towards the right of the paper.
    }
    \label{exfig:Optical tweezers}
\end{figure}

\begin{figure}[htbp]
    \centering
    \includegraphics{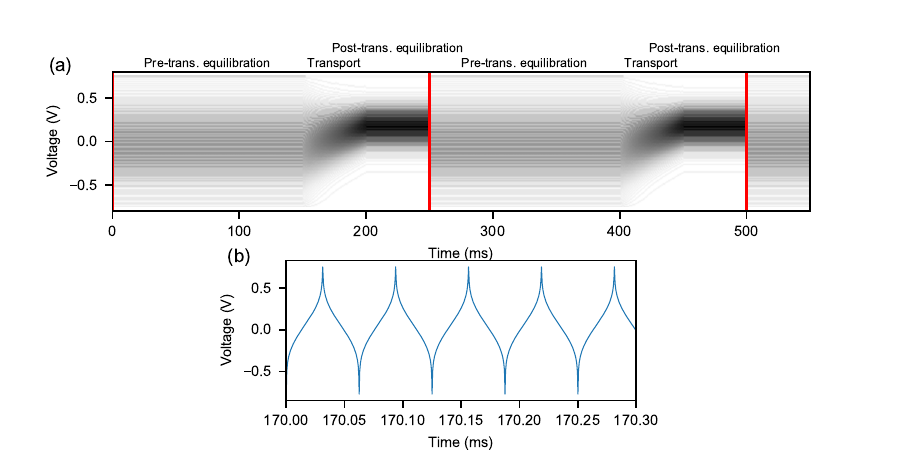}
    \caption{{\bf Exemplified scan pattern for the translation and compression protocol with $\tau=\SI{50}{ms}$}. The vertical axis is the voltage input for the EOD device (the same magnitude of voltages are applied to the two EOD devices), which diagonally translates the laser in the x-y plane to extend the scan range.
    The duration between red lines corresponds to a cycle, which was repeated more than 3,000 times for $\tau\le\SI{200}{ms}$ and more than 1,500 times for $\tau=\SI{500}{ms}$ for each particle. 
    (b) is the magnification of (a).
    In this protocol, we generated Gaussian-profile potentials with a width of approximately \SI{500}{nm}.
    This width is sufficiently large so that the particle effectively feels a harmonic potential (see the first row in Extended Data Fig.~\ref{exfig:potential}).}
    \label{exfig:ScanPattern:translation}
\end{figure}

\begin{figure}[htbp]
    \centering
    \includegraphics{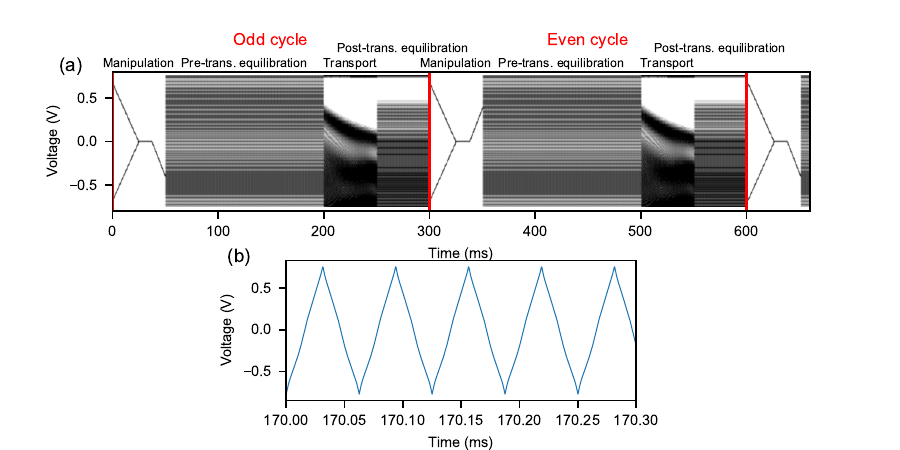}
    \caption{{\bf Exemplified scan pattern for the information erasure with $\tau=\SI{50}{ms}$}. See the caption of External Data Fig.~\ref{exfig:ScanPattern:translation} for details.
    The manipulation step is intended to balance the initial condition between \textbf{0} and \textbf{1}. The initial value of the equilibration step is forced to be \textbf{0} or \textbf{1} alternatively.
    We repeated this cycle more than 5,000 times for each particle.
    }
    \label{exfig:ScanPattern:landauer}
\end{figure}

\begin{figure}[htbp]
    \centering
    \includegraphics{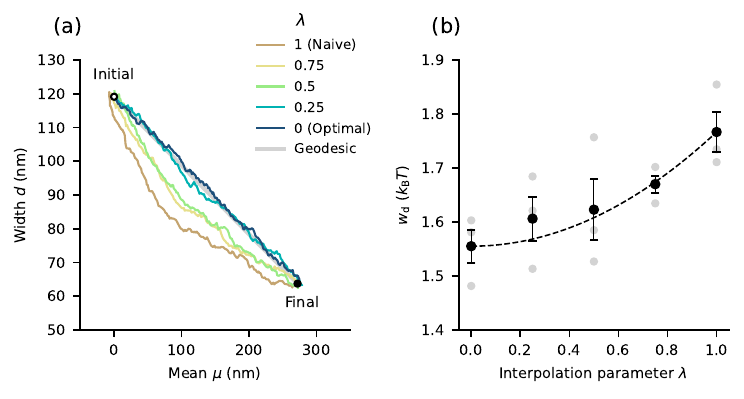}
    \caption{{\bf Transports with intermediate optimality}. We linearly interpolated the transport between naive and optimal transports with the interpolation parameter $\lambda$. $\tau = \SI{50}{ms}$. See SI Section~\ref{SI:intermediate optimality} for the details. (a) Trajectories in the $(\mu, d)$ space.   (b) Dependence of $\Wdis$ on $\lambda$. The experimental data coincided with the theory (dashed curve) given by Eq.~\eqref{eq:non-optimality}, validating the theory within error bars.  The gray symbol corresponds to an experimental run consisting of more than 3,000 repetitions for a particle.
 We performed three runs with three independent particles under each condition to measure the mean values (black symbols). The error bars correspond to s.e.m. (three samples each). 
    }
    \label{exfig:nonoptimality}
\end{figure}

\begin{figure}[htbp]
    \centering
    \includegraphics{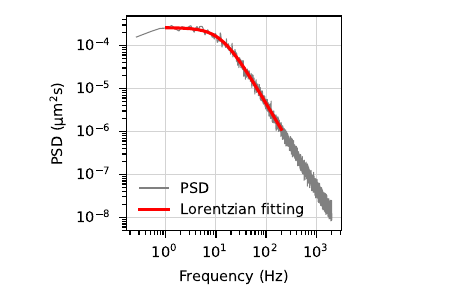}
    \caption{{\bf Power spectrum density (PSD) of the particle position.}
    The PSD is measured for a particle trapped in a harmonic potential created by a scanning laser. The typical trajectory length is five minutes.
    The red curve is a fitting by a Lorentzian function Eq.~\eqref{eq:Lorentzian}; $C(f)= \frac{\kBT}{2\pi^2\gamma}\cdot\frac 1{f^2+f_0^2}$, where $f$ is the frequency, and $f_0$ is the corner frequency. We use $fC(f)$ for the fitting so that the fitting weight is biased to the region close to the corner frequency $f_0$.
    See Methods for the details.
    }
    \label{exfig:power spectrum}
\end{figure}

\begin{figure}[htbp]
    \centering
    \includegraphics{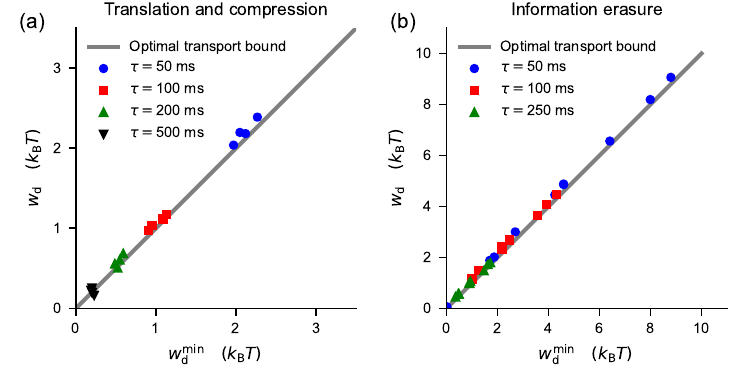}
    \caption{{\bf Comparison between $\Wdis$ and $\Wdis^\m$ for each run}.
    Here, a run consists of thousands of repetitions for a choice of particle. We did four independent runs with four particles in the translation and compression and three independent runs with three particles in the information erasure under each condition to measure the mean values in Figs.~\ref{fig:translation}e, \ref{fig:landauer}d, and \ref{fig:tradeoff}. (a) The translation and compression protocol. (b) The information erasure protocol.
    The solid line corresponds to the bound given by the optimal transport (Eq.~\eqref{eq:Wdis:min}).
    A few points are slightly below this line due to statistical variations.
    }
    \label{exfig:Wdis:each}
\end{figure}

\begin{figure*}[bth]
    \centering
    \includegraphics{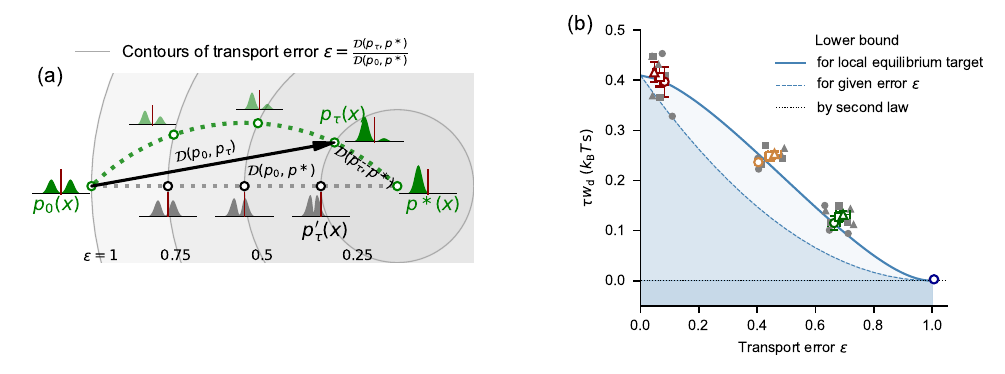}
    \caption{{\bf Trade-off between energy cost, speed, and accuracy in terms of the Wasserstein distance.}
        (a) Schematic of the geometry in the distribution space.
        The black arrow indicates the optimal transport to a given final distribution $p_\tau(x)$.
        The green distribution series illustrates the target distributions in the present experiments, which are designed so that they are approximately local equilibrium in each well of a double well potential.
        $\pideal(x)$ is the distribution with perfect erasure.
        The transport error $\varepsilon$ is defined based on the Wasserstein distance between $p_\tau(x)$ and $\pideal(x)$ (Eq.~\eqref{eq:transport error}).
        The gray distribution series are on the geodesic between $p_0(x)$ and $\pideal(x)$. $\Wdis$ is minimized with the optimal transport to these distributions for given $\varepsilon$. 
        $p_\tau^\prime(x)$ illustrates a distribution with the same $\varepsilon$ as a given distribution $p_\tau(x)$ but on the geodesic.
    (b) The dissipated work $\Wdis$ was scaled by multiplying $\tau$ since $\Wdis$ is scaled by $1/\tau$ for optimal transport (Eq.~\eqref{eq:Wdis:min}).
    The three curves correspond to the lower bounds for given $p_0$ and $p_\tau$ with local equilibrium target (solid, Eq.~\eqref{eq:trade-off:1}), for given $\varepsilon$ (dashed, Eq.~\eqref{eq:trade-off:2}), and by second law (dotted, Eq.~\eqref{eq:trade-off:second low}).
     The target distribution with perfect erasure $\pideal(x)$ was constructed using the mean final distributions $p_\tau(x)$ of $\alpha=1$ as $\pideal(x)=p_\tau(x) /[\int_{-\infty}^0 dx'p_\tau(x')]$ for $x<0$ and $\pideal(x) = 0$ for $x\ge 0$.
     We mixed $\pideal(x)$ and $p_0(x)$ to prepare a series of distributions to draw the solid curve. 
     See Extended Data Fig.~\ref{exfig:tradeoff:dist} for details.
     The colors and symbols are the same as those in Fig.~\ref{fig:landauer}d.
The error bars indicate s.e.m. (three samples each).
    }
    \label{exfig:tradeoff}
\end{figure*}

\begin{figure}[htbp]
    \centering
    \includegraphics{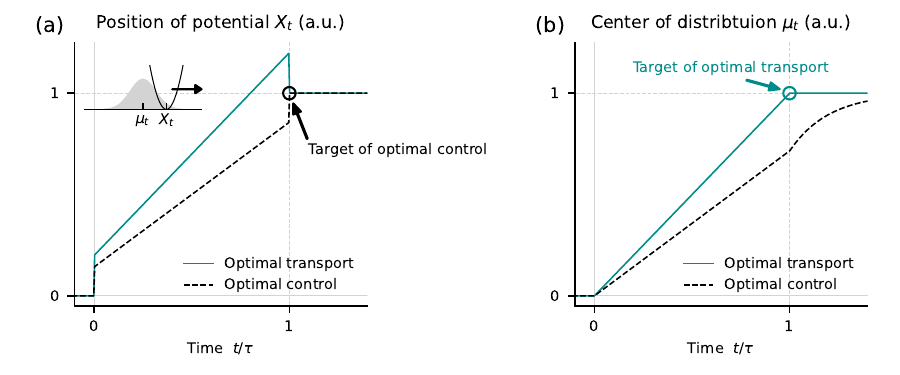}
    \caption{{\bf Quantitative comparison between optimal transport and optimal control.} 
The difference between those protocols is illustrated for a simple translation protocol without compression.
We assumed a Brownian particle obeying the Langevin equation with the friction coefficient $\gamma$ and a harmonic potential $k(x-X_t)^2/2$.
$k$ is a fixed spring constant, and $X_t$ is the position of the potential.
The ratio $k\tau/\gamma$ corresponds to the ratio of protocol time to relaxation time and is set to be 5, which solely determines the optimal dynamics.
The variation in this ratio does not alter the qualitative behavior of the dynamics.
In this situation with harmonic potentials with a fixed $k$, the probability distribution is always Gaussian with a constant width.
Its center is denoted as $\mu_t$.
Theoretical values\cite{Nakazato2021, Schmiedl2007} are plotted; time course of the position of the potential $X_t$ for the two protocols (a), and time course of the center of the distribution $\mu_t$ (b).
The task of the optimal transport is to reach a given final distribution, which is a Gaussian with the center at $\mu_\tau=1$, at $t=\tau$ with the minimum dissipation (b, blue circle).
The task of the optimal control is to change a set of control parameters of the potentials to the final target values $X_\tau=1$ at $t=\tau$ with the minimum dissipation (a, black circle).
In the optimal control case, the distribution does not reach the desired final distribution (i.e., the equilibrium state of the final potential) in finite time, while relaxing to it in a sufficiently long time after $\tau$  (b, black solid curve).
Thus, optimal control does not care about the final state at $t=\tau$.
    See SI section~\ref{SI:optimal control} for details.
    }
    \label{exfig:OptimalControl}
\end{figure}

\begin{figure}[htbp]
    \centering
    \includegraphics{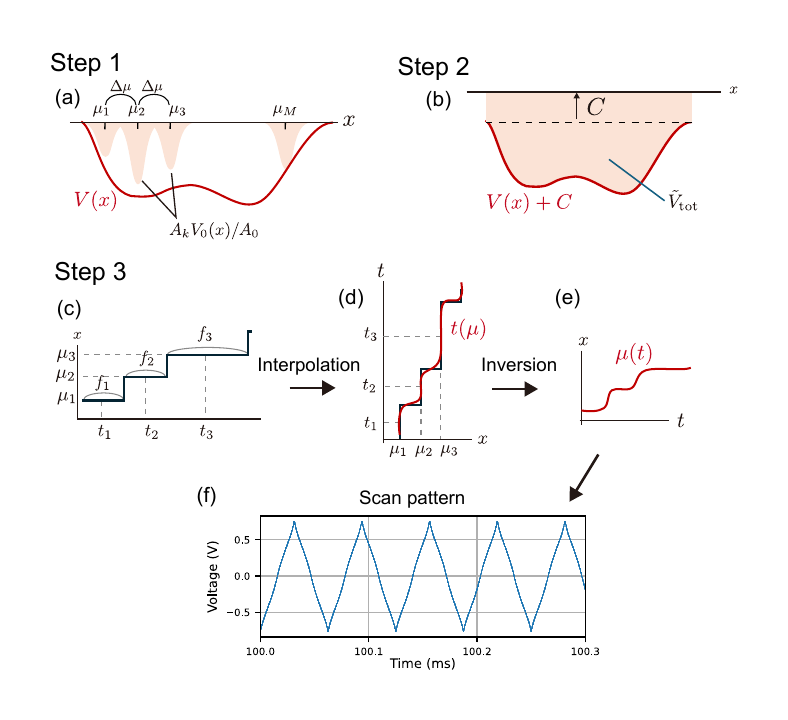}
    \caption{{\bf Schematic of the generation of the laser scan pattern.} We fit the target potential $V(x)$ by mixed Gaussian distributions (Step1), shift the baseline to adjust the integral of potential (Step 2), and generate a scan pattern (Step 3).
    See SI Section~\ref{SI:Scanning optical tweezers} for the details.
    }
    \label{exfig:Scan}
\end{figure}

\begin{figure}[htbp]
    \centering
    \includegraphics{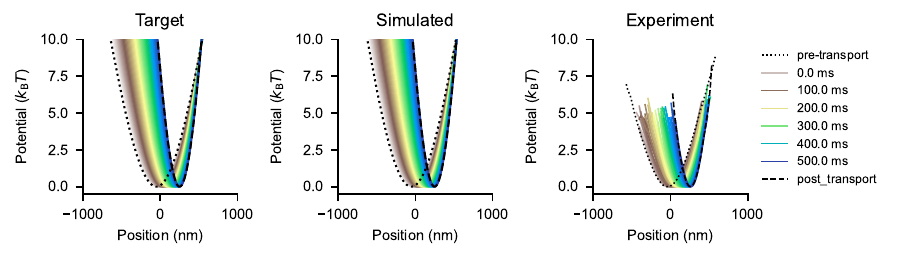}
    \includegraphics{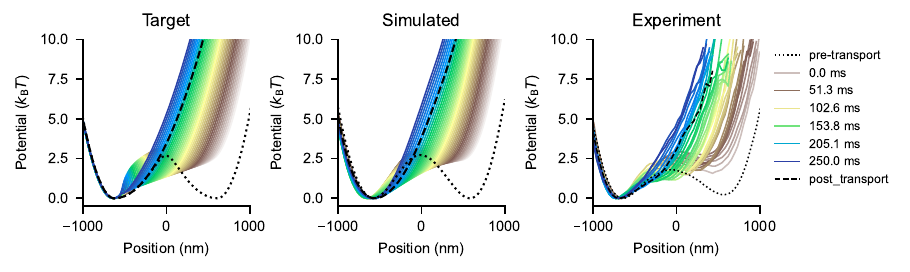}
    \includegraphics{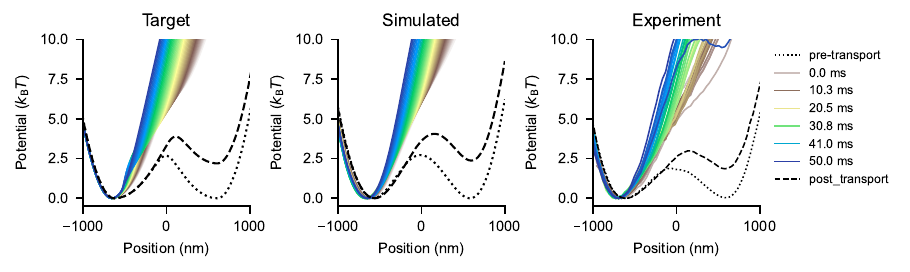}
    \includegraphics{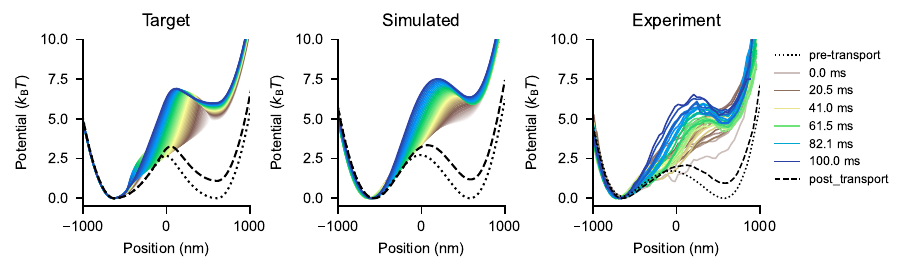}
    \caption{{\bf Recovery of the potential profiles from the experimental trajectories.} Typical examples of translation and compression (first row, optimal protocol with $\tau=\SI{500}{ms}$) and information erasure ($\alpha = 1$ and $\tau = \SI{250}{ms}$, $\alpha = 0.9$ and $\tau = \SI{50}{ms}$, and $\alpha = 0.75$ and $\tau = \SI{100}{ms}$, respectively, from second to fourth rows).
    {\it Target} potentials are the optimal potential profiles (see Methods and SI Section~\ref{SI:Optimal protocol} for the details).
    In the experiments, multiple factors, including the finite laser spot size and finite update rates of the scan patterns, smoothen the potential profiles, which are simulated in {\it Simulated} potentials.
    The potentials recovered from experimental trajectories are {\it Experiment} potentials.
    We used multiple frames during the transport. The method developed here (see Methods for the details) based on the drift velocity is applicable to nonequilibrium trajectories.
    Colors indicate different frames.
    Dotted and dashed curves are the potentials in the pre- and post-transport steps, respectively.
    }
    \label{exfig:potential}
\end{figure}

\begin{figure}[htbp]
    \centering
    \includegraphics{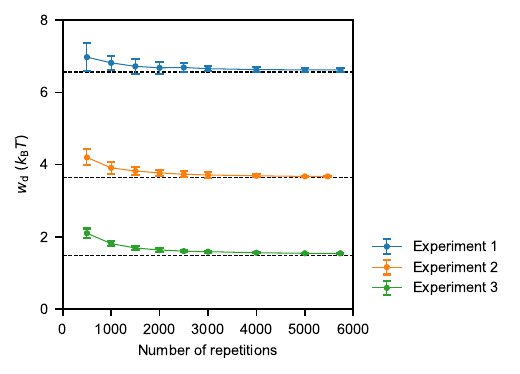}
    \caption{{\bf Dependence of $\Wdis$ evaluated from the distributions on the number of repetitions.} See Methods for the details.
    We randomly sampled repetitions with specified numbers and evaluated $\Wdis$ from the distributions.
    We repeated this 50 times and calculated the mean and s.d.
    Different colors correspond to different experimental runs (Experiments 1, 2, and 3 correspond to 50-, 100-, and 250-ms Landauer transport with $\alpha=1$, respectively). 
    The dashed lines indicate the values of $\Wdis$ evaluated using the recovered potentials based on Eq.~\eqref{eq:W, F}.
    The plots show that more than 3,000 repetitions are necessary for sufficient convergence in this setup (information erasure), but it should depend on the transport protocols.
    }
    \label{exfig:Methods:Wdis:cycleN}
\end{figure}

\begin{figure}[htbp]
    \centering
    \includegraphics{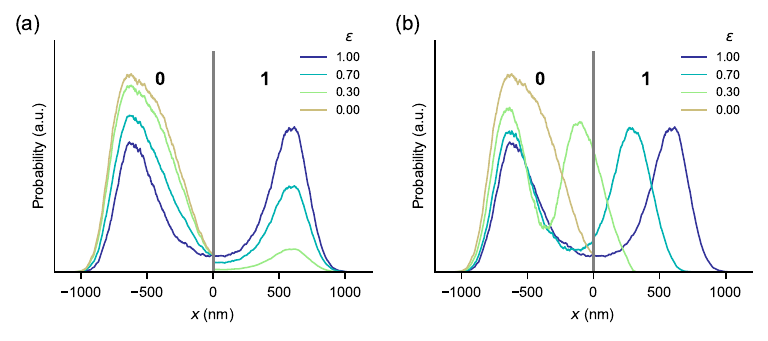}
    \caption{{\bf Exemplified probability density for generating the trade-off curves in Extended Data Fig.~\ref{exfig:tradeoff}b.}
    (a) Distributions constructed by mixing $p_0(x)$ and $\pideal(x)$ as $\beta p_0(x)+(1-\beta)\pideal(x)$ with a mixing parameter $\beta$. These are used for generating data to draw the solid curve in Extended Data Fig.~\ref{exfig:tradeoff}b.
    (b) Distributions constructed by optimal transport between $p_0(x)$ and $\pideal(x)$, which are located on the geodesic in the distribution space between $p_0$ and $\pideal(x)$ (Extended Data Fig.~\ref{exfig:tradeoff}a), for the dashed curve in Extended Data Fig.~\ref{exfig:tradeoff}b.
    The mean distribution of nine distributions, including different $\tau$ values, with $\alpha=1$ at $t=0$ is used as $p_0(x)$.
    The region in $x<0$ of the mean distribution at $t=\tau$ with $\alpha=1$ is trimmed and normalized to create $\pideal(x)$.
    The distributions with $\varepsilon=1$ and 0 correspond to $p_0(x)$ and $\pideal(x)$, respectively.
    }
    \label{exfig:tradeoff:dist}
\end{figure}

\clearpage

\end{bibunit}

\clearpage
\noindent\textsf{\textbf{\Large Supplementary information for
``Experimentally achieving minimal dissipation via thermodynamically optimal transport''}}

\noindent
\textsf{\textbf{Shingo Oikawa, Yohei Nakayama, Sosuke Ito, Takahiro Sagawa, and Shoichi Toyabe}}

\begin{bibunit}
\renewcommand{\bibnumfmt}[1]{S#1.}
\renewcommand{\citenumfont}[1]{S#1}

\renewcommand{\theequation}{S\arabic{equation}}
\setcounter{equation}{0}

\setcounter{section}{0}  
\setcounter{subsection}{0} 
\setcounter{subsubsection}{0} 

\def\thesection{S\arabic{section}}
\def\thesubsection{S\arabic{section}.\arabic{subsection}}
\makeatletter
\def\p@subsection{}
\def\p@subsubsection{}
\makeatother
\def\thesubsubsection{S\arabic{section}.\arabic{subsection}.\arabic{subsubsection}}

\section{Generation of laser scan pattern}
\label{SI:Scanning optical tweezers}

The laser focal point was scanned by an electro-optical deflector at \SI{16}{kHz} to create a trapping potential.
The translation speed was adjusted so that the mean light intensity at each position is proportional to the target potential at each position.
We deconvolved target potential profiles by Gaussian intensity profile that approximates the laser spot to obtain the scan pattern under constraints that the total power is fixed, the spatial scanning range is limited, and the mean time duration residing at each position is positive.

\newcommand{\Itot}{I_\mathrm{tot}}
\newcommand{\Utot}{V_\mathrm{tot}}
\newcommand{\Vtot}{V_\mathrm{tot}}
\newcommand{\Lmax}{{L_\mathrm{max}}}
\newcommand{\Lmin}{{L_\mathrm{min}}}

Let the intensity profile of a single laser spot be $I_0(x)$, which could be approximated by a Gaussian profile.
The particle feels a potential proportional to $I_0(x)$ at each location $x$.
Think that we scan the laser spot at a rate faster than the relaxation time of the particle, which is typically the ratio of the trapping spring constant and the friction coefficient of the particle (estimated to be in the order of \SI{100}{Hz} in the present setup).
Then, the particle feels only the mean light intensity at each position $x$ given by
\begin{align}\label{eq:I(x)}
    I(x)=\int^\infty_{-\infty}I_0(x-X)f(X)\dd X
\end{align}
for a fixed scanning protocol.
Here, $X$ is the laser position, and $f(X)$ is the residential time fraction at $X$.
We modulate only the scan speed at each position and do not temporally modulate $I_0(x)$.
Therefore, the total light intensity 
$\Itot=\int_{-\infty}^{\infty}I(x)\dd x$ is constant.
The particle feels a potential at each position proportional to $I(x)$, $V(x)=-aI(x)$.
$a$ is a proportional coefficient.
Since $\Itot$ is a constant independent of the scanning protocol, 
$\Utot=\int_{-\infty}^{\infty}V(x)\dd x=-a\Itot$ is also a constant.
We obtain a similar expression to Eq.~\eqref{eq:I(x)},
\begin{align}\label{eq:U(x)}
    V(x)=\int^\infty_{-\infty}V_0(x-X)f(X)\dd X.
\end{align}
$V_0(x)=-aI_0(x)$ is the potential profile of a single laser spot.

\(\Utot\) is inferred by experiments as follows.
We generate a Gaussian potential $V(x)$ with a known value of the width $\sigma$.
$V(x)$ should has the form of \(V(x) = \frac{\Utot}{\sqrt{2\pi}\sigma}\exp\left(-\frac{x^2}{2\sigma^2}\right)\) with an unknown parameter $\Utot$, which will be determined by experiments.
Since the particle is strongly trapped, we approximate $V(x)$ as a harmonic potential $\kappa x^2/2$ with the effective spring constant $\kappa$.
We measure $\kappa$ from the width of the equilibrium distribution $p(x)$ using $p(x) \propto \exp(-\kappa x^2/(2\kBT))$.
Finally, we obtain $\Utot$ as \(\Utot = -\sqrt{2\pi}\sigma^3 \kappa\).

The actual workflow of the deconvolution and the generation of the scan pattern proceeds in the following three steps (Extended Data Fig.~\ref{exfig:Scan}).
\begin{description}
\item[Step 1.] The target potential $V(x)$ is approximated by the superposition of $M$ elementary potentials located in the scan range between 
$L_\mathrm{min}=\SI{-1.17}{\micro m}$ and $L_\mathrm{max}=\SI{1.17}{\micro m}$ (Extended Data Fig.~\ref{exfig:Scan}a).
\begin{align}
    \tilde V(x)=\sum^M_{k=1}\frac{A_k}{A_0}V_0(x-\mu_k),\qquad \mu_k=\Lmin+(k-1)\Delta \mu.
\end{align}
We used $M=40$ in the experiments.
We assume that the laser spots are located at $\mu_k$ separated by $\Delta \mu=(\Lmax-\Lmin)/(M-1)$, and each potential has a Gaussian profile $V_0(x)=\frac {A_0}{\sqrt{2\pi\sigma_0^2}}\exp\left(-\frac{x^2}{2\sigma_0^2}\right)$.
$\sigma_0$ is the spot size of the laser.
To evaluate $\sigma_0$, we generated a potential by scanning the laser at a high frequency between two locations in a distance of $s$.
The particle feels a double-well potential for $s>\sigma_0$ and a single-well potential for $s\le\sigma_0$.
We measured $\sigma_0$ to be \SI{260}{nm} as the threshold distance of $s$ by controlling $s$ and measuring the position distribution of the trapped particle. 

We determine $\{A_k\}$ by minimizing a cost function defined as
\begin{align}
    S(\{A_k\})=\int^{\Lmax}_{\Lmin}[V(x)-\tilde V(x)]^2\dd x+b_1\left[\sum_{k=1}^{M-1}(A_{k+1}-A_k)^2\right]^{b_2}.
\end{align}
The second term on the right-hand side is introduced so that neighboring peaks do not have large amplitude differences, which may cause distortions in the scan patterns in the experiments.
$b_1$ and $b_2$ are constants and adjusted depending on the target potential profile.
We used $b_1=0.01$ and $b_2 = 0.7$.
We find $\{A_k\}$ that minimize $S(\{A_k\})$ under the constraint of $A_k<0,\,\forall k$ by trust region reflective algorithm of SciPy library \cite{2020SciPy-NMeth}.
\item[Step 2.]  
The total of the potential
\begin{align}
    \tilde V_\mathrm{tot}=\int^\infty_{-\infty}\sum^M_{k=1}\frac{A_k}{A_0}V_0(x-\mu_k)\dd x=\sqrt{2\pi}\sigma_0\sum^M_{k=1}A_k
\end{align}
should be a constant, $V_\mathrm{tot}$, which is determined by an experiment, as explained.
To satisfy this constraint, we introduce an offset $C$ such that
\begin{align}
    C =\frac{V_\mathrm{tot}-\tilde V_\mathrm{tot}}{\Lmax-\Lmin}\cdot\frac{\tilde V_\mathrm{tot, in}}{\tilde V_\mathrm{tot}},
\end{align}
and modify the cost function as
\begin{align}
    S(\{A_k\}; C) =\int^{\Lmax}_{\Lmin}[V(x)+C-\tilde V(x)]^2\dd x+b_1\left[\sum_{k=1}^{M-1}(A_{k+1}-A_k)^2\right]^{b_2}.
\end{align}
Here, $\tilde V_\mathrm{tot, in}=\int^\Lmax_\Lmin\tilde V(x)\dd x$ is the total of potential within $[\Lmin,\Lmax]$.
We iterate the update of $C$ and minimum finding of $S$ multiple times (21 times) to find the optimal values of $\{A_k\}$ under the constraint of $A_k<0,\,\forall k$ (Extended Data Fig.~\ref{exfig:Scan}b).

\item[Step 3.]  We determine the laser scan pattern based on normalized peak intensities $f_k=A_k/\sum^M_{k=1} A_k$, which correspond to the residential time fraction of the laser spot at $\mu_k$ during the scan cycle (Extended Data Fig.~\ref{exfig:Scan}c). The center of the duration is $t_m=\sum^m_{k=1}f_k-\frac{f_m}2$ for $1\le m\le M$, $t_0=0$, and $t_{M+1}$ = 1.
Accordingly, the spot locations are extended by adding $\mu_0=\Lmin$ and $\mu_{M+1}=\Lmax$ for the interpolation in the next step.
$\{(t_m,\mu_m)\}$ is interpolated to obtain a continuous function $t(\mu)$ (Extended Data Fig.~\ref{exfig:Scan}d). The inversed function $\mu(t)$ (Extended Data Fig.~\ref{exfig:Scan}e) and $\mu(1-t)$ are alternatively repeated to generate the scan pattern (Extended Data Fig.~\ref{exfig:Scan}f).
\end{description}

The scan pattern was controlled by a computer equipped with LabVIEW and a multifunction card (NI, TX).
The voltage update rate for EOD was \SI{3.3}{MHz}.
A diagonal direction of the two-dimensional EOD control was used to obtain the maximum travel distance.

To calibrate the laser power variation with position, we measured the spring constants of a harmonic potential generated by laser spots at fixed positions by fitting the distributions with Gaussians.
The spring constants are measured 10 times each at 21 different locations for \SI{20}{s} at each location.
The dependence of the spring constant on the position was fitted by an eighth-order polynomial function.


\section{Optimal transport}
\label{SI:Optimal protocol}

\subsection{Computation of optimal transport}

We used the optimal transport map to calculate the distribution dynamics under the optimal transport protocol \(p_t^\opt\).
The optimal transport protocol in the one-dimensional Euclidean space translates each segment in the initial distribution $p_0$ to the final distribution $p_\tau$ linearly without changing the positional order of the segments\cite{Villani2009} (Fig.~\ref{fig:intro}c, right).
The optimal transport map \(\mathcal{T}_{p_0\to p_\tau}(x)\) describes where a segment at \(x\) in \(p_0\) is translated.
The optimal transport map is related with \({\cal A}_{p\to q}(x)\) as 
\(\mathcal{T}_{p\to q}(x)={\operatorname{argmin}}_{{\cal A}_{p\to q}}\int\dd x ||x-{\cal A}_{p\to q}(x)||^2p(x)\).
The preservation of the positional order of the segments means
\begin{align}\label{eq:equality:cumulative-distribution-function}
 \Cdf_\tau(\mathcal{T}_{p_0\to p_\tau}(x)) = \Cdf_0(x),
\end{align}
where \(\Cdf_0(x) = \int_{-\infty}^x \dd x' p_0(x')\) and \(\Cdf_\tau(x) = \int_{-\infty}^x \dd x' p_\tau(x')\) are the cumulative distribution function at \(t=0\) and \(t=\tau\), respectively.
Since the translation is linear, a segment at \(x\) in \(p_0\) is translated to the position \(x_t^\opt(x) = \left(1 - \frac{t}{\tau}\right) x + \frac{t}{\tau}\mathcal{T}_{p_0\to p_\tau}(x)\) in \(p_t\).
That is, the optimal cumulative distribution function \(\Cdf_t^\opt (x) =\int_{-\infty}^x {\rm d} x' p^\opt_t(x')\) satisfies
\begin{align}\label{eq:equality:cumulative-distribution-function:t}
\Cdf_t^\opt\left(x_t^\opt(x)\right) = \Cdf_0(x).
\end{align}
Therefore, we first obtain \(\mathcal{T}_{p_0\to p_\tau}\) by numerically solving Eq.~(\ref{eq:equality:cumulative-distribution-function}) and then calculate \(p_t^\opt\) by numerically differentiating \(\Cdf_t^\opt\) obtained from Eq.~(\ref{eq:equality:cumulative-distribution-function:t}).

The details of the calculation are as follows.
To obtain \(\mathcal{T}_{p_0\to p_\tau}\), we find pairs of positions \((y_i, z_i)\) satisfying
\begin{align}\label{eq:separated equality}
 \Cdf_0(y_i) &= \cdf_i, & \Cdf_\tau(z_i) &= \cdf_i,
\end{align} by the bisection method.
Here, \(\cdf_i\) are chosen as
\begin{align}
 \cdf_i =
 \begin{cases}
  \cdf_\mathrm{lower} + 2^{(i-M_\mathrm{log}-1)} \Delta\cdf
  & i = 0, \cdots, M_\mathrm{log} - 1
  \\
  \cdf_\mathrm{lower} + \left(i - M_\mathrm{log} + \frac{1}{2}\right) \Delta\cdf
  & i = M_\mathrm{log}, \cdots, M_\mathrm{log} + M_\mathrm{linear} - 1
  \\
  \cdf_\mathrm{upper} - 2^{-(i-M_\mathrm{linear}-M_\mathrm{log}+2)} \Delta\cdf
  & i = M_\mathrm{linear} + M_\mathrm{log}, \cdots, M_\mathrm{linear} + 2M_\mathrm{log} - 1
 \end{cases}
 ,
\end{align}
where \(\cdf_\mathrm{lower} = 0\), \(\cdf_\mathrm{upper} = 1\), \(M_\mathrm{linear} = 65536\), \(\Delta\cdf = \left(\cdf_\mathrm{upper} - \cdf_\mathrm{lower}\right) / M_\mathrm{linear}\), and \(M_\mathrm{log} = 32\).
Since \(\Cdf_\tau(z_i) = \Cdf_\tau\left(\mathcal{T}_{p_0\to p_\tau}(y_i)\right)\) follows from Eq.~(\ref{eq:equality:cumulative-distribution-function}) and Eq.~(\ref{eq:separated equality}), \(z_i\) equal to \(\mathcal{T}_{p_0\to p_\tau}(y_i)\).
Therefore, \(\Cdf_t^\opt\left( \left(1-\frac{t}{\tau}\right) y_i + \frac{t}{\tau}z_i\right)\) is evalutaed as \(\phi_i\).
We obtain \(p_t^\opt\) at \(x = \left(1-\frac{t}{\tau}\right) \frac{y_i + y_{i+1}}{2} + \frac{t}{\tau} \frac{z_i+z_{i+1}}{2}\) by numerically differentiating \(\Cdf_t^\opt\) as 
\begin{align}
 p_t^\opt\left(x\right) = \frac{\phi_{i+1} - \phi_i}{\left[\left(1-\frac{t}{\tau}\right) y_{i+1} + \frac{t}{\tau}z_{i+1}\right] - \left[\left(1-\frac{t}{\tau}\right) y_i + \frac{t}{\tau}z_i\right]}
 ,
 \label{eq:differentiation of cumulative}
\end{align}
and interpolate it as necessary.

\subsection{Computation of potential dynamics}
\label{SI:Computation of potential}

Once distribution dynamics $p_t$ is obtained, we can calculate the potential dynamics $V_t$ that realizes $p_t$ by solving a Fokker-Planck equation,
\begin{align}\label{eq:Fokker-Planck}
    \frac{\partial p_t(x)}{\partial t}=-\frac{\partial}{\partial x}\left[-\frac{1}{\gamma}\frac{\partial V_t(x)}{\partial x}p_t(x) -\frac{\kBT}{\gamma}\frac{\partial p_t(x)}{\partial x}\right]
    ,
\end{align}
with boundary conditions \(J_t(x) = 0\) at \(x\to\pm\infty\).
Here,
\begin{align}
    J_t(x) = -\frac{1}{\gamma}\frac{\partial V_t(x)}{\partial x}p_t(x) -\frac{\kBT}{\gamma}\frac{\partial p_t(x)}{\partial x}
    \label{eq:flux}
\end{align}
is the probability flux.
We solve Eq.~(\ref{eq:Fokker-Planck}) by splitting it into
\begin{align}
    \frac{\partial p_t(x)}{\partial t} &= -\frac{\partial J_t(x)}{\partial x}
    ,
    \label{eq:continuous}
    \\
    \frac{1}{p_t(x)}\left[\gamma J_t(x) + \kBT\frac{\partial p_t(x)}{\partial x}\right] &= -\frac{\partial V_t(x)}{\partial x}
    .
    \label{eq:flux transformed}
\end{align}
Eq.~(\ref{eq:flux transformed}) is obtained from Eq.~(\ref{eq:flux}).
We first obtain \(J_t(x)\) by integrating Eq.~(\ref{eq:continuous}), and then calculate \(V_t(x)\) by integrating Eq.~(\ref{eq:flux transformed}).
For Gaussian dynamics, as in translation and compression protocols, we can derive simplified equations (Eq.~\eqref{eq:ODE:mu,sigma}).

To numerically integrate Eqs.~(\ref{eq:continuous}) and (\ref{eq:flux transformed}), 
we consider a sufficiently wide interval \([x_\mathrm{min}, x_\mathrm{max}]\) and impose boundary conditions \(J_t(x) = 0\) at \(x_\mathrm{min} = \SI{-1.17}{\micro m}\) and \(x_\mathrm{max} = \SI{1.17}{\micro m}\).
We use \(p_{t_i}(x_j)\) at \(t_i = i \Delta t\) and \(x_j = x_\mathrm{min} + \left(j+\frac{1}{2}\right)\Delta x\) to calculate \(J_t\) and \(V_t\), where \(0 \leq i \leq N_t\) and \(0 \leq j \leq N_x\) are integers, \(\Delta t = \tau/N_t\), \(\Delta x = (x_\mathrm{max} - x_\mathrm{min})/(N_x + 1)\), \(N_t = 40\), and \(N_x = 200\).
\(J_{t_i+\Delta t/2}(x_k)\) is obtained by summing up the discretized version of the left hand side of Eq.~(\ref{eq:continuous}) multiplied by \(\Delta x\), \(\left[p_{t_{i+1}}(x_j) - p_{t_i}(x_j)\right]\frac{\Delta x}{\Delta t}\), where \(0 < k < N_x\) is a half-integer.
To reduce the effect of numerical error, we calculate \(J^\rightarrow_{t_i+\Delta t / 2}(x_k)\) and \(J^\leftarrow_{t_i+\Delta t / 2}(x_k)\) by summing up from \(x_\mathrm{min}\) and \(x_\mathrm{max}\), respectively.
And then, we adopt 
\(J^\rightarrow_{t_i+\Delta t / 2}(x_k)\) for \(x_k < x^*\),
\(J^\leftarrow_{t_i+\Delta t / 2}(x_k)\) for \(x_k > x^*\),
and \([J^\rightarrow_{t_i+\Delta t / 2}(x_k) + J^\leftarrow_{t_i+\Delta t / 2}(x_k)]/2\) for \(x_k = x^*\) as \(J_{t_i+\Delta t / 2}(x_k)\),
where we choose
\begin{align}
    x^* = \mathop{\mathrm{argmin}}_{x_k} \left|\frac{J^\rightarrow_{t_i+\Delta t / 2}(x_k) - J^\leftarrow_{t_i+\Delta t / 2}(x_k)}{J^\rightarrow_{t_i+\Delta t / 2}(x_k) + J^\leftarrow_{t_i+\Delta t / 2}(x_k)}\right|
    .
\end{align}
Finally, we discretize the left hand side of Eq.~(\ref{eq:flux transformed}) as
\begin{align}
    \frac{
    \gamma J_{t_i+\Delta t / 2}(x_j + \Delta x / 2) + \kBT \frac{1}{2\Delta x} \left[p_{t_i}(x_{j+1}) - p_{t_i}(x_{j}) + p_{t_{i+1}}(x_{j+1}) - p_{t_{i+1}}(x_{j})\right]
    }
    {\frac{1}{4}\left[p_{t_i}(x_{j+1}) + p_{t_i}(x_{j}) + p_{t_{i+1}}(x_{j+1}) + p_{t_{i+1}}(x_{j})\right]}
    ,
\end{align}
and obtain \(V_{t_i+\Delta t / 2}(x_j)\) by summing up it.

\subsection{Translation and compression protocols}

For Gaussian dynamics, the mean $\mu_t$ and width $d_t$ of the distribution obey
\begin{align}\label{eq:ODE:mu,sigma}
    \frac{\dd\mu_t}{\dd t}=-\frac {K_t}\gamma(\mu_t-X_t),\qquad \frac{\dd d_t^2}{\dd t}=-\frac{2K_t}\gamma\left(d_t^2-\frac{\kBT}{K_t}\right),
\end{align}
where \(X_t\) and \(K_t\) are the position and the stiffness of a harmonic potential, respectively.
Equation~\eqref{eq:ODE:mu,sigma} are deduced from the Fokker-Planck equation (Eq.~\eqref{eq:Fokker-Planck}),
and leads to
\begin{align}\label{eq:KX}
    K_t=\frac 1{d_t^2}\left(\kBT-\frac\gamma 2\frac{\dd d_t^2}{\dd t}\right),\qquad
    X_t=\mu_t+\frac\gamma{K_t}\frac{\dd\mu_t}{\dd t}.
\end{align}
These relations provide $K_t$ and $X_t$ for given $\mu_t$ and $d_t$.

The {\it optimal} transport corresponds to a linear translation of $\mu_t$ and $d_t$ between the initial and final distributions:
\begin{align}
    \mu^\opt_t=(\mu_\tau-\mu_0)\frac t\tau+\mu_0,\quad
    d^\opt_t=(d_\tau-d_0)\frac t\tau+d_0.
\end{align}

The {\it gearshift} transport is a combination of two optimal transports with different translation speeds and time duration:
\begin{align}
    \mu^\gearshift_t&=\frac rs(\mu_\tau-\mu_0)\frac t\tau+\mu_0,\quad
    d^\gearshift_t=\frac rs(d_\tau-d_0)\frac t\tau+d_0, &\mathrm{for}\quad t\le s\tau,\\
    \mu^\gearshift_t&=\frac{1-r}{1-s}(\mu_{\tau}-\mu_0)\frac {t-s\tau}\tau+\mu^\gearshift_{s\tau},\quad
    d^\gearshift_t=\frac{1-r}{1-s}(d_{\tau}-d_0)\frac {t-s\tau}\tau+d^\gearshift_{s\tau}, &\mathrm{for}\quad t>s\tau
    .
\end{align}
$r=1/3$ and $s=2/3$ are the fractions of the distance and duration before the gearshift, respectively.
We obtain $K_t$ and $X_t$ for the optimal and gearshift transports by using Eq.~\eqref{eq:KX}.

The {\it naive} protocol linearly varies the position $X_t$ and stiffness $K_t$ of the potential as
\begin{align}
K^\naive_t=& (K_\tau-K_0)\frac t\tau+K_0, \quad
X^\naive_t=(X_\tau-X_0)\frac t\tau+X_0.
\end{align}

In the experiment, we generated Gaussian-profile potentials with a width of approximately \SI{500}{nm}.
This width is sufficiently large so that the particle effectively feels a harmonic potential (Extended Data Fig.~\ref{exfig:potential}).

\subsection{Transports with intermediate optimality}
\label{SI:intermediate optimality}

The intermediate transports between the optimal and naive transports are obtained by simple interpolation:
\begin{align}
    \mu^\lambda_t = \lambda\mu_t^\naive + (1-\lambda)\mu_t^\opt,\quad 
    d^\lambda_t = \lambda d_t^\naive + (1-\lambda)d_t^\opt.
\end{align}
Here, $\lambda$ is the interpolation parameter. $\lambda =0$ and 1 correspond to the optimal and naive protocols, respectively.
Given that the naive and optimal transports have the same initial and final distributions, $\mu_0^\opt=\mu_0^\naive$, $d_0^\opt=d_0^\naive$, $\mu_\tau^\opt=\mu_\tau^\naive$, $d_\tau^\opt=d_\tau^\naive$, the intermediate transport also has the same initial and final distributions.

Then, we can show that the dissipated work of this intermediate transport $\Wdis^\lambda$ depends quadratically on \(\lambda\) as
\begin{align}\label{eq:non-optimality}
    \Wdis^\lambda = \lambda^2\Wdis^\naive+(1-\lambda^2)\Wdis^\opt=\Wdis^\opt+\lambda^2\left(\Wdis^\naive-\Wdis^\opt\right).
\end{align}
Equation~\eqref{eq:non-optimality} is obtained as follows.
In Gaussian processes, the dissipated work for ($\mu_t, d_t$)  is given by 
\begin{align}
\Wdis=\gamma \int_0^\tau\left(\frac{\dd\mathcal{L}}{\dd t}\right)^2\dd t.
\end{align}
where the square of the speed in distribution space $(\frac{{\rm d}\mathcal{L}}{{\rm d} t})^2$ is defined as
\begin{align}\label{eq:dLdt}
    \left(\frac{\dd\mathcal{L}}{\dd t}\right)^2 = \left(\frac{\dd\mu_t}{\dd t}\right)^2 + \left(\frac{\dd d_t}{\dd t}\right)^2.
\end{align}
To consider the dissipated work for ($\mu_t^{\lambda}, d_t^{\lambda}$), we calculate $\int_0^\tau \left(\frac{\dd\mu^\lambda_t}{\dd t}\right)^2 \dd t$ as
\begin{align}
    \int_0^\tau\left(\frac{\dd\mu^\lambda_t}{\dd t}\right)^2\dd t = \lambda^2\int_0^\tau\left(\frac{\dd\mu^\naive_t}{\dd t}\right)^2\dd t+2\lambda(1-\lambda)\int_0^\tau\frac{\dd\mu^\naive_t}{\dd t}\frac{\dd\mu^\opt_t}{\dd t}\dd t+(1-\lambda)^2\int_0^\tau\left(\frac{\dd\mu^\opt_t}{\dd t}\right)^2\dd t.
\end{align}
Since $\frac{\dd\mu^\opt_t}{\dd t}=\frac{\mu_\tau-\mu_0}\tau$ is independent of $t$,
\begin{align}
    \int_0^\tau\frac{\dd\mu^\naive_t}{\dd t}\frac{\dd\mu^\opt_t}{\dd t}\dd t = \frac{\mu_\tau-\mu_0}\tau\int_0^\tau\frac{\dd\mu^\naive_t}{\dd t}\dd t=\frac{(\mu_\tau-\mu_0)^2}\tau=\int_0^\tau\left(\frac{\dd\mu^\opt_t}{\dd t}\right)^2\dd t.
\end{align}
Therefore, 
\begin{align}
    \int_0^\tau\left(\frac{\dd\mu^\lambda_t}{\dd t}\right)^2\dd t = \lambda^2\int_0^\tau\left(\frac{\dd\mu^\naive_t}{\dd t}\right)^2dt+(1-\lambda^2)\int_0^\tau\left(\frac{\dd\mu^\opt_t}{\dd t}\right)^2\dd t.
\end{align}
From the same calculation for $d^\lambda_t$, we calculate $\int_0^\tau\left(\frac{\dd d^\lambda_t}{\dd t}\right)^2\dd t $ as  
\begin{align}
\int_0^\tau\left(\frac{\dd d^\lambda_t}{\dd t}\right)^2\dd t = \lambda^2\int_0^\tau\left(\frac{\dd d^\naive_t}{\dd t}\right)^2dt+(1-\lambda^2)\int_0^\tau\left(\frac{\dd d^\opt_t}{\dd t}\right)^2\dd t,
\end{align}
and thus we obtain Eq.~\eqref{eq:non-optimality} by using the above equations and Eq.~\eqref{eq:dLdt} for ($\mu_t^{\lambda}, d_t^{\lambda}$), ($\mu_t^{\rm naive}, d_t^{\rm naive}$) and ($\mu_t^{\rm opt}, d_t^{\rm opt}$).

We experimentally validated Eq.~\eqref{eq:non-optimality} by linearly interpolating transports between naive and optimal transports (Extended Data Fig.~\ref{exfig:nonoptimality}), which shows the quadratic dependence of $\Wdis$ on $\lambda$.

\subsection{Information erasure}
\label{SI:Information erasure}

The initial and final distributions, $p_0(x)$ and $p_\tau(x)$, are designed by combining Gaussians with different widths:
\begin{align}
p(x;\alpha)=\alpha g(x+m) + (1-\alpha)g(-x+m),
\end{align}
where
\begin{align}
g(x)= {\displaystyle \frac{1}{\sqrt{\pi/2}(s_\mathrm{L}+s_\mathrm{R})}e^{-x^2/2s_\mathrm{L}^2}}\quad (x<0),\qquad
{\displaystyle g(x)= \frac{1}{\sqrt{\pi/2}(s_\mathrm{L}+s_\mathrm{R})}e^{-x^2/2s_\mathrm{R}^2}}\quad (x\ge 0).
\end{align}
We used the peak position $m=\SI{600}{nm}$, and the widths $s_\mathrm{L}=\SI{120}{nm}$ and $s_\mathrm{R}=\SI{240}{nm}$.
We used $\alpha=0.5$ for the initial distribution.
We varied $\alpha$ of the final distribution to control the accuracy ($\alpha = 0.5, 0.75, 0.9$, and 1).

\section{Energy-speed-accuracy trade-off}
\label{SI:varepsilon}

\subsection{Energy-speed-accuracy trade-off in terms of Wasserstein distance}

In this paper, the accuracy of information erasure is quantified by $\eta_\tau$, which is the fraction of \textbf{0} at $t=\tau$ (Figs.~\ref{fig:landauer}c and \ref{fig:tradeoff}).
In Ref.\cite{Klinger2025}, the transport error is defined based on the Wasserstein distance.
We apply this framework to the present experimental setup of the information erasure.

We define transport error $\varepsilon$ by
\begin{align}
    \varepsilon = \frac{\w(p_\tau, \pideal)}{\w(p_0, \pideal)},
    \label{eq:transport error}
\end{align}
according to Ref.\cite{Klinger2025}.
Here, $\pideal$ is a target distribution.
We chose $\pideal(x)=p_\tau(x) /[\int_{-\infty}^0 dx'p_\tau(x')]$ for $x<0$ and $\pideal(x) = 0$ for $x\ge 0$, which corresponds to a perfect information erasure.

We can derive hierarchical trade-off relations between energy cost, speed, and accuracy\cite{Klinger2025}.
Let $p_\tau'(x)$ be the distribution with the same $\varepsilon$ as $p_\tau(x)$ and on the geodesic between $p_0(x)$ and $\pideal(x)$ (Extended Data Fig.~\ref{exfig:tradeoff}a).
That is, \(p_\tau'(x)\) satisfies \(\w(p_0, \pideal) = \w(p_0, p_\tau') + \w(p_\tau', \pideal)\).
On the one hand, a triangle inequality $\w(p_0, \pideal) \le \w(p_0, p_\tau)+\w(p_\tau, \pideal)$ and $\w(p_\tau,\pideal)=\w(p_\tau',\pideal)$ lead to $\w(p_0, p_\tau') \le \w(p_0, p_\tau)$.
On the other hand, 
$1-\varepsilon=1-\w(p_\tau, \pideal)/\w(p_0, \pideal)
    =1-\w(p_\tau', \pideal)/\w(p_0, \pideal)
    = [\w(p_0, \pideal)-\w(p_\tau', \pideal)]/\w(p_0, \pideal) 
    =\w(p_0,p_\tau')/\w(p_0, \pideal)$.
Using these relations, hierarchical trade-off relations are derived:
\begin{align}
\tau\cdot\Wdis\cdot (1-\varepsilon)^{-2} &\ge \gamma \w(p_0,p_\tau)^2\cdot (1-\varepsilon)^{-2} \label{eq:trade-off:1}\\
&\ge \gamma \w(p_0,\pideal)^2  \label{eq:trade-off:2}\\  
&\ge 0. \label{eq:trade-off:second low}
\end{align}

The first inequality (Eq.~\eqref{eq:trade-off:1}) corresponds to a bound of $\Wdis$ for given $p_0$ and $p_\tau$ (solid line in Extended Data Fig.~\ref{exfig:tradeoff}b), which is achieved in the experiments (symbols).
The equality is satisfied by the optimal transport, which is also indicated as solid lines in Fig.~\ref{fig:landauer}d and \ref{fig:tradeoff}.

On the other hand, we can find other distributions that achieve smaller $\Wdis$ with the same $\varepsilon$.
Equation~\eqref{eq:trade-off:2} gives the minimum of $\Wdis$ for given $\varepsilon$  (dashed line in Extended Data Fig.~\ref{exfig:tradeoff}b).
Since $\gamma\w(p_0,\pideal)^2$ is a constant for given $\gamma$ and $p_0$, the second inequality indicates the trade-off where smaller $\varepsilon$ and $\tau$ require larger $\Wdis$ even if we change the choice of \(p_\tau\).
While this bound is achieved when $p_\tau(x)$ lies on the geodesic between $p_0(x)$ and $\pideal(x)$, such a distribution has a rather complicated profile and would not be practical for information processing (gray distribution series in Extended Data Fig.~\ref{exfig:tradeoff}a, see also Extended Data Fig.~\ref{exfig:tradeoff:dist}b).
In fact, if we finalize the transport protocol by a double-well potential, it costs additional dissipation for the relaxation to local equilibrium in each well \cite{Proesmans2020, Proesmans_2020-2}. 
As noted, the present transport targets approximately local-equilibrium distributions in each well of a double-well potential (illustrated as green distribution series in Extended Data Fig.~\ref{exfig:tradeoff}a). 

The fact that $\tau\Wdis$ is bound by a finite value (Eq.~\eqref{eq:trade-off:2}) means that the equality in Eq.~\eqref{eq:trade-off:second low}, $\tau\Wdis=0$, is not achievable even in the quasi-static limit $\tau\to\infty$,  while the second-law bound, $\Wdis=0$, is achievable in $\tau\to\infty$.
This is because $\Wdis$ vanishes with the order of $1/\tau$ at $\tau\to\infty$, implying that $\tau\Wdis$ remains at a finite value.

\subsection{Relation between the transport error and accuracy}

The above results demonstrate that, while $\eta_\tau$ is also a measure to quantify the accuracy of the transport, the Wasserstein distance provides a unified perspective on the trade-off by quantifying both the error and energy cost.
We derive that $\varepsilon$ is related to $\eta_\tau$ as $\varepsilon^2 \propto 1-\eta_\tau$ under
the assumption that \(p_\tau\) is written as \(p_\tau(x) = \pideal(x) + \delta \psi(x)\) with a small parameter \(\delta\), and \(\pideal(x) > 0\) holds for \(x < 0\).

We evaluate the optimal transport map \(\mathcal{T}_{p_\tau\to \pideal}\) to the leading order in \(\delta\).
\(\mathcal{T}_{p_\tau\to \pideal}\) satisfies
\begin{align}
 \Cdf_\tau(x) = \Cdf^*\left(\mathcal{T}_{p_\tau\to \pideal}(x)\right),
 \label{eq:equality:cumulative-distribution-function:pq}
\end{align}
where \(\Cdf_\tau\) and \(\Cdf^*\) are the cumulative distribution functions given as
\(\Cdf_\tau(x) = \int_{-\infty}^x \dd x' p_\tau(x')\)
and
\(\Cdf^*(x) = \int_{-\infty}^x \dd x' \pideal(x')\), respectively.
For \(x>0\), \(\Cdf_\tau\) is evaluated as
\begin{align}
 \Cdf_\tau(x)
 = 1 - \int_x^\infty \dd x' p_\tau(x')
 = 1 - \delta\int_x^\infty \dd x' \psi(x')
 = 1 + O(\delta)
 .
 \label{eq:estimation-of-Phi}
\end{align}
Combining Eqs.~(\ref{eq:equality:cumulative-distribution-function:pq}) and (\ref{eq:estimation-of-Phi}) with \(\Cdf^*(0) = 1\), we obtain \(\Cdf^*\left(\mathcal{T}_{p_\tau\to \pideal}(x)\right) - \Cdf^*(0) = O(\delta)\), or equivalently \(-\int_{\mathcal{T}_{p_{\tau} \to p^* }(x)}^0 dx' p^*(x')= O(\delta)\). Here, $\mathcal{T}_{p_{\tau} \to p^* }(x)<0$ since $p^*(x) =0$ for $x\geq0$.
Because \(\pideal(x)\) is nonzero for $x<0$ and independent of \(\delta\), \(\mathcal{T}_{p_\tau\to \pideal}(x) = O(\delta)\) holds for \(x>0\).
On the other hand, for \(x<0\), \(\Cdf_\tau\) is evaluated as
\begin{align}
 \Cdf_\tau(x)
 = \int_{-\infty}^x \dd x' p_\tau(x')
 = \Cdf^*(x) + \delta \Psi(x)
 ,
\end{align}
where \(\Psi(x) = \int_{-\infty}^x \dd x' \psi(x')\). From Eq.~(\ref{eq:equality:cumulative-distribution-function:pq}), we obtain 
 \(\Cdf^*\left(\mathcal{T}_{p_\tau\to \pideal}(x)\right) - \Cdf^*(x) = O(\delta)\) for $x<0$. Thus, \(\int_{\mathcal{T}_{p_{\tau} \to p^* }(x)}^x dx' p^*(x')= O(\delta)\)  and
 \(\mathcal{T}_{p_\tau\to \pideal}(x) - x = O(\delta)\) hold for $x<0$.

From the above results, we obtain
\begin{align}
 \w(p_\tau, \pideal)^2 =& \int_{-\infty}^\infty \dd x \Vert x - \mathcal{T}_{p_\tau\to \pideal}(x)\Vert^2 p_\tau(x)
 \nonumber \\
 =& \int_{-\infty}^0 \dd x \Vert O(\delta)\Vert^2 p_\tau(x)
 + \int_0^\infty \dd x \Vert x - O(\delta)\Vert^2 p_\tau(x)
 = \delta \int_0^\infty \dd x \Vert x \Vert^2 \psi(x) + O(\delta^2)
 ,
\end{align}
where we used $p_{\tau}(x) =\delta \psi(x)$ for $x>0$.
As a result, \(\varepsilon^2\) is written as
\begin{align}
 \varepsilon^2 = c \delta \int_0^\infty \dd x \psi(x) + O(\delta^2) = c (1 - \eta_\tau) + O(\delta^2),
\end{align}
where
\begin{align}
 c = \frac{1}{\w(p_0, \pideal)^2}\frac{\int_0^\infty \dd x \Vert x \Vert^2 \psi(x)}{\int_0^\infty \dd x \psi(x)}
\end{align}
is a coefficient independent of \(\delta\). Here, we used $1- \eta_{\tau} = \int_0^{\infty} dx p_{\tau}(x) = \delta \int dx \psi_{\tau} (x)$, which is obtained from $p_{\tau}(x) =\delta \psi(x)$ for $x>0$ and $\int_{-\infty}^{\infty} p_{\tau}(x)=1$.
Therefore, \(\varepsilon^2\) is  proportional to \(1 - \eta_\tau\) up to the order \(O(\delta )\).

\section{Comparision between optimal transport and optimal control}
\label{SI:optimal control}

A similar but fundamentally distinct approach to minimization of dissipative work is the concept of {\it optimal control}, which seeks the optimal protocol for given initial and final control parameters instead of probability distributions.

A quantitative difference between optimal transport and optimal control is illustrated in Extended Data Fig.~\ref{exfig:OptimalControl} for a simple translation protocol without compression, which translates a harmonic potential with a fixed spring constant.
In such a situation, the distribution is always Gaussian with a constant width.
Let $X_t$ be the position of a harmonic potential, and $\mu_t$ be the mean particle position.

The task of the optimal transport is to reach a given final distribution of the particle position at $t=\tau$, which is a Gaussian with the center at $\mu_\tau$, with the minimum dissipation.
The optimal dynamics of $X_t$ has a forward jump at $t=0$ and a backward jump at $t=\tau$ (Extended Data Fig.~\ref{exfig:OptimalControl}a, blue curve).
This results in a linear translation of $\mu_t$, which reaches the given final position at $t=\tau$ (Extended Data Fig.~\ref{exfig:OptimalControl}b, blue curve).

On the other hand, the task of optimal control is to change a set of control parameters of the potentials to the final target values at $t=\tau$ with the minimum dissipation.
In this situation, $X_t$ also has initial and final jumps, but both are in the forward direction, different from the optimal transport (Extended Data Fig.~\ref{exfig:OptimalControl}a, black dashed curve).
This protocol translates the distribution linearly but does not reach the equilibrium distribution of the final target potential within finite time (Extended Data Fig.~\ref{exfig:OptimalControl}b, black dashed curve).
The distribution relaxes to the equilibrium one in a sufficiently long time after $\tau$.
Thus, optimal control does not necessarily produce the desired final distribution in finite time.

Therefore, optimal transport and optimal control are based on fundamentally different strategies for different purposes.
Optimal transport is more relevant for processes like information erasure, where one wants to produce the given final state \textbf{0} in finite time.


\end{bibunit}


\begin{thebibliography}{10}
\urlstyle{rm}
\expandafter\ifx\csname url\endcsname\relax
  \def\url#1{\texttt{#1}}\fi
\expandafter\ifx\csname urlprefix\endcsname\relax\def\urlprefix{URL }\fi
\expandafter\ifx\csname doiprefix\endcsname\relax\def\doiprefix{DOI: }\fi
\providecommand{\bibinfo}[2]{#2}
\providecommand{\eprint}[2][]{\url{#2}}

\bibitem{Monge1781}
\bibinfo{author}{Monge, G.}
\newblock \emph{\bibinfo{title}{M{\'e}moire sur la th{\'e}orie des d{\'e}blais
  et des remblais}} (\bibinfo{publisher}{De l'Imprimerie Royale},
  \bibinfo{year}{1781}).

\bibitem{Villani2009}
\bibinfo{author}{Villani, C.}
\newblock \emph{\bibinfo{title}{Optimal transport: old and new}}
  (\bibinfo{publisher}{Springer}, \bibinfo{year}{2009}).

\bibitem{Arjovsky2017}
\bibinfo{author}{Arjovsky, M.}, \bibinfo{author}{Chintala, S.} \&
  \bibinfo{author}{Bottou, L.}
\newblock \bibinfo{title}{Wasserstein generative adversarial networks.}
\newblock In \bibinfo{editor}{Precup, D.} \& \bibinfo{editor}{Teh, Y.~W.}
  (eds.) \emph{\bibinfo{booktitle}{Proceedings of the 34th International
  Conference on Machine Learning (ICML 2017)}}, vol.~\bibinfo{volume}{70},
  \bibinfo{pages}{214--223} (\bibinfo{publisher}{PMLR}, \bibinfo{year}{2017}).

\bibitem{lipman2023flow}
\bibinfo{author}{Lipman, Y.}, \bibinfo{author}{Chen, R. T.~Q.},
  \bibinfo{author}{Ben-Hamu, H.}, \bibinfo{author}{Nickel, M.} \&
  \bibinfo{author}{Le, M.}
\newblock \bibinfo{title}{Flow matching for generative modeling}.
\newblock In \emph{\bibinfo{booktitle}{The Eleventh International Conference on
  Learning Representations}} (\bibinfo{year}{2023}).

\bibitem{Schiebinger2019}
\bibinfo{author}{Schiebinger, G.} \emph{et~al.}
\newblock \bibinfo{journal}{\bibinfo{title}{Optimal-transport analysis of
  single-cell gene expression identifies developmental trajectories in
  reprogramming}}.
\newblock {\emph{\JournalTitle{Cell}}} \textbf{\bibinfo{volume}{176}},
  \bibinfo{pages}{928--943.e22}, \doiprefix\url{10.1016/j.cell.2019.01.006}
  (\bibinfo{year}{2019}).

\bibitem{Aurell2012}
\bibinfo{author}{Aurell, E.}, \bibinfo{author}{Gaw{\c e}dzki, K.},
  \bibinfo{author}{Mejía-Monasterio, C.}, \bibinfo{author}{Mohayaee, R.} \&
  \bibinfo{author}{Muratore-Ginanneschi, P.}
\newblock \bibinfo{journal}{\bibinfo{title}{Refined second law of
  thermodynamics for fast random processes}}.
\newblock {\emph{\JournalTitle{J. Stat. Phys.}}}
  \textbf{\bibinfo{volume}{147}}, \bibinfo{pages}{487–505},
  \doiprefix\url{10.1007/s10955-012-0478-x} (\bibinfo{year}{2012}).

\bibitem{Nakazato2021}
\bibinfo{author}{Nakazato, M.} \& \bibinfo{author}{Ito, S.}
\newblock \bibinfo{journal}{\bibinfo{title}{Geometrical aspects of entropy
  production in stochastic thermodynamics based on {Wasserstein} distance}}.
\newblock {\emph{\JournalTitle{Phys. Rev. Res.}}} \textbf{\bibinfo{volume}{3}},
  \bibinfo{pages}{043093}, \doiprefix\url{10.1103/physrevresearch.3.043093}
  (\bibinfo{year}{2021}).

\bibitem{Proesmans2020}
\bibinfo{author}{Proesmans, K.}, \bibinfo{author}{Ehrich, J.} \&
  \bibinfo{author}{Bechhoefer, J.}
\newblock \bibinfo{journal}{\bibinfo{title}{Finite-time {Landauer} principle}}.
\newblock {\emph{\JournalTitle{Phys. Rev. Lett.}}}
  \textbf{\bibinfo{volume}{125}}, \bibinfo{pages}{100602},
  \doiprefix\url{10.1103/physrevlett.125.100602} (\bibinfo{year}{2020}).

\bibitem{Proesmans_2020-2}
\bibinfo{author}{Proesmans, K.}, \bibinfo{author}{Ehrich, J.} \&
  \bibinfo{author}{Bechhoefer, J.}
\newblock \bibinfo{journal}{\bibinfo{title}{Optimal finite-time bit erasure
  under full control}}.
\newblock {\emph{\JournalTitle{Phys. Rev. E}}} \textbf{\bibinfo{volume}{102}},
  \bibinfo{pages}{032105}, \doiprefix\url{10.1103/physreve.102.032105}
  (\bibinfo{year}{2020}).

\bibitem{Landauer1961}
\bibinfo{author}{Landauer, R.}
\newblock \bibinfo{journal}{\bibinfo{title}{Irreversibility and heat generation
  in the computing process}}.
\newblock {\emph{\JournalTitle{IBM J. Res. Dev.}}}
  \textbf{\bibinfo{volume}{5}}, \bibinfo{pages}{183–191},
  \doiprefix\url{10.1147/rd.53.0183} (\bibinfo{year}{1961}).

\bibitem{Parrondo2015}
\bibinfo{author}{Parrondo, J. M.~R.}, \bibinfo{author}{Horowitz, J.~M.} \&
  \bibinfo{author}{Sagawa, T.}
\newblock \bibinfo{journal}{\bibinfo{title}{Thermodynamics of information}}.
\newblock {\emph{\JournalTitle{Nature Phys.}}} \textbf{\bibinfo{volume}{11}},
  \bibinfo{pages}{131–139}, \doiprefix\url{10.1038/nphys3230}
  (\bibinfo{year}{2015}).

\bibitem{Benamou2000}
\bibinfo{author}{Benamou, J.-D.} \& \bibinfo{author}{Brenier, Y.}
\newblock \bibinfo{journal}{\bibinfo{title}{A computational fluid mechanics
  solution to the {Monge--Kantorovich} mass transfer problem}}.
\newblock {\emph{\JournalTitle{Num. Math.}}} \textbf{\bibinfo{volume}{84}},
  \bibinfo{pages}{375–393}, \doiprefix\url{10.1007/s002110050002}
  (\bibinfo{year}{2000}).

\bibitem{Aurell2011}
\bibinfo{author}{Aurell, E.}, \bibinfo{author}{Mejía-Monasterio, C.} \&
  \bibinfo{author}{Muratore-Ginanneschi, P.}
\newblock \bibinfo{journal}{\bibinfo{title}{Optimal protocols and optimal
  transport in stochastic thermodynamics}}.
\newblock {\emph{\JournalTitle{Physical Review Letters}}}
  \textbf{\bibinfo{volume}{106}}, \bibinfo{pages}{250601},
  \doiprefix\url{10.1103/physrevlett.106.250601} (\bibinfo{year}{2011}).

\bibitem{chen2019stochastic}
\bibinfo{author}{Chen, Y.}, \bibinfo{author}{Georgiou, T.~T.} \&
  \bibinfo{author}{Tannenbaum, A.}
\newblock \bibinfo{journal}{\bibinfo{title}{Stochastic control and
  nonequilibrium thermodynamics: Fundamental limits}}.
\newblock {\emph{\JournalTitle{IEEE Transactions on Automatic Control}}}
  \textbf{\bibinfo{volume}{65}}, \bibinfo{pages}{2979--2991},
  \doiprefix\url{10.1109/TAC.2019.2939625} (\bibinfo{year}{2019}).

\bibitem{ito2024geometric}
\bibinfo{author}{Ito, S.}
\newblock \bibinfo{journal}{\bibinfo{title}{Geometric thermodynamics for the
  {Fokker}--{Planck} equation: stochastic thermodynamic links between
  information geometry and optimal transport}}.
\newblock {\emph{\JournalTitle{Information Geometry}}}
  \textbf{\bibinfo{volume}{7}}, \bibinfo{pages}{441--483},
  \doiprefix\url{10.1007/s41884-023-00102-3} (\bibinfo{year}{2024}).

\bibitem{Seifert2012}
\bibinfo{author}{Seifert, U.}
\newblock \bibinfo{journal}{\bibinfo{title}{Stochastic thermodynamics,
  fluctuation theorems and molecular machines}}.
\newblock {\emph{\JournalTitle{Rep. Prog. Phys.}}}
  \textbf{\bibinfo{volume}{75}}, \bibinfo{pages}{126001},
  \doiprefix\url{10.1088/0034-4885/75/12/126001} (\bibinfo{year}{2012}).

\bibitem{Ciliberto2017}
\bibinfo{author}{Ciliberto, S.}
\newblock \bibinfo{journal}{\bibinfo{title}{Experiments in stochastic
  thermodynamics: Short history and perspectives}}.
\newblock {\emph{\JournalTitle{Phys. Rev. X}}} \textbf{\bibinfo{volume}{7}},
  \bibinfo{pages}{021051}, \doiprefix\url{10.1103/physrevx.7.021051}
  (\bibinfo{year}{2017}).

\bibitem{Pigolotti-Peliti}
\bibinfo{author}{Peliti, L.} \& \bibinfo{author}{Pigolotti, S.}
\newblock \emph{\bibinfo{title}{Stochastic Thermodynamics: An Introduction}}
  (\bibinfo{publisher}{Princeton University Press}, \bibinfo{year}{2021}).

\bibitem{Shiraishi2016}
\bibinfo{author}{Shiraishi, N.}, \bibinfo{author}{Saito, K.} \&
  \bibinfo{author}{Tasaki, H.}
\newblock \bibinfo{journal}{\bibinfo{title}{Universal trade-off relation
  between power and efficiency for heat engines}}.
\newblock {\emph{\JournalTitle{Phys. Rev. Lett.}}}
  \textbf{\bibinfo{volume}{117}}, \bibinfo{pages}{190601},
  \doiprefix\url{10.1103/physrevlett.117.190601} (\bibinfo{year}{2016}).

\bibitem{Sagawa2009}
\bibinfo{author}{Sagawa, T.} \& \bibinfo{author}{Ueda, M.}
\newblock \bibinfo{journal}{\bibinfo{title}{Minimal energy cost for
  thermodynamic information processing: Measurement and information erasure}}.
\newblock {\emph{\JournalTitle{Phys. Rev. Lett.}}}
  \textbf{\bibinfo{volume}{102}}, \bibinfo{pages}{250602},
  \doiprefix\url{10.1103/physrevlett.102.250602} (\bibinfo{year}{2009}).

\bibitem{Lutz2015}
\bibinfo{author}{Lutz, E.} \& \bibinfo{author}{Ciliberto, S.}
\newblock \bibinfo{journal}{\bibinfo{title}{Information: From {Maxwell}’s
  demon to {Landauer}’s eraser}}.
\newblock {\emph{\JournalTitle{Phys. Today}}} \textbf{\bibinfo{volume}{68}},
  \bibinfo{pages}{30–35}, \doiprefix\url{10.1063/pt.3.2912}
  (\bibinfo{year}{2015}).

\bibitem{Berut2012}
\bibinfo{author}{Bérut, A.} \emph{et~al.}
\newblock \bibinfo{journal}{\bibinfo{title}{Experimental verification of
  {Landauer}’s principle linking information and thermodynamics}}.
\newblock {\emph{\JournalTitle{Nature}}} \textbf{\bibinfo{volume}{483}},
  \bibinfo{pages}{187–189}, \doiprefix\url{10.1038/nature10872}
  (\bibinfo{year}{2012}).

\bibitem{Jun2014}
\bibinfo{author}{Jun, Y.}, \bibinfo{author}{Gavrilov, M.} \&
  \bibinfo{author}{Bechhoefer, J.}
\newblock \bibinfo{journal}{\bibinfo{title}{High-precision test of
  {Landauer’s} principle in a feedback trap}}.
\newblock {\emph{\JournalTitle{Phys. Rev. Lett.}}}
  \textbf{\bibinfo{volume}{113}}, \bibinfo{pages}{190601},
  \doiprefix\url{10.1103/physrevlett.113.190601} (\bibinfo{year}{2014}).

\bibitem{Gavrilov2016}
\bibinfo{author}{Gavrilov, M.} \& \bibinfo{author}{Bechhoefer, J.}
\newblock \bibinfo{journal}{\bibinfo{title}{Erasure without work in an
  asymmetric double-well potential}}.
\newblock {\emph{\JournalTitle{Phys. Rev. Lett.}}}
  \textbf{\bibinfo{volume}{117}}, \bibinfo{pages}{200601},
  \doiprefix\url{10.1103/physrevlett.117.200601} (\bibinfo{year}{2016}).

\bibitem{RibezziCrivellari2019}
\bibinfo{author}{Ribezzi-Crivellari, M.} \& \bibinfo{author}{Ritort, F.}
\newblock \bibinfo{journal}{\bibinfo{title}{Large work extraction and the
  {Landauer} limit in a continuous {Maxwell} demon}}.
\newblock {\emph{\JournalTitle{Nature Phys.}}} \textbf{\bibinfo{volume}{15}},
  \bibinfo{pages}{660–664}, \doiprefix\url{10.1038/s41567-019-0481-0}
  (\bibinfo{year}{2019}).

\bibitem{Dago2021}
\bibinfo{author}{Dago, S.}, \bibinfo{author}{Pereda, J.},
  \bibinfo{author}{Barros, N.}, \bibinfo{author}{Ciliberto, S.} \&
  \bibinfo{author}{Bellon, L.}
\newblock \bibinfo{journal}{\bibinfo{title}{Information and thermodynamics:
  Fast and precise approach to {Landauer}’s bound in an underdamped
  micromechanical oscillator}}.
\newblock {\emph{\JournalTitle{Phys. Rev. Lett.}}}
  \textbf{\bibinfo{volume}{126}}, \bibinfo{pages}{170601},
  \doiprefix\url{10.1103/physrevlett.126.170601} (\bibinfo{year}{2021}).

\bibitem{Dago2023}
\bibinfo{author}{Dago, S.}, \bibinfo{author}{Ciliberto, S.} \&
  \bibinfo{author}{Bellon, L.}
\newblock \bibinfo{journal}{\bibinfo{title}{Adiabatic computing for optimal
  thermodynamic efficiency of information processing}}.
\newblock {\emph{\JournalTitle{Proc. Nat. Acad. Sci.}}}
  \textbf{\bibinfo{volume}{120}}, \bibinfo{pages}{e2301742120},
  \doiprefix\url{10.1073/pnas.2301742120} (\bibinfo{year}{2023}).

\bibitem{Toyabe2010}
\bibinfo{author}{Toyabe, S.}, \bibinfo{author}{Sagawa, T.},
  \bibinfo{author}{Ueda, M.}, \bibinfo{author}{Muneyuki, E.} \&
  \bibinfo{author}{Sano, M.}
\newblock \bibinfo{journal}{\bibinfo{title}{Experimental demonstration of
  information-to-energy conversion and validation of the generalized
  {Jarzynski} equality}}.
\newblock {\emph{\JournalTitle{Nature Phys.}}} \textbf{\bibinfo{volume}{6}},
  \bibinfo{pages}{988–992}, \doiprefix\url{10.1038/nphys1821}
  (\bibinfo{year}{2010}).

\bibitem{Koski2014}
\bibinfo{author}{Koski, J.~V.}, \bibinfo{author}{Maisi, V.~F.},
  \bibinfo{author}{Pekola, J.~P.} \& \bibinfo{author}{Averin, D.~V.}
\newblock \bibinfo{journal}{\bibinfo{title}{Experimental realization of a
  {Szilard} engine with a single electron}}.
\newblock {\emph{\JournalTitle{Proc. Nat. Acad. Sci.}}}
  \textbf{\bibinfo{volume}{111}}, \bibinfo{pages}{13786–13789},
  \doiprefix\url{10.1073/pnas.1406966111} (\bibinfo{year}{2014}).

\bibitem{Toyabe2010PRL}
\bibinfo{author}{Toyabe, S.} \emph{et~al.}
\newblock \bibinfo{journal}{\bibinfo{title}{Nonequilibrium energetics of a
  single {F$_1$-ATPase} molecule}}.
\newblock {\emph{\JournalTitle{Phys. Rev. Lett.}}}
  \textbf{\bibinfo{volume}{104}}, \bibinfo{pages}{198103},
  \doiprefix\url{10.1103/physrevlett.104.198103} (\bibinfo{year}{2010}).

\bibitem{DiTerlizzi2024}
\bibinfo{author}{Di~Terlizzi, I.} \emph{et~al.}
\newblock \bibinfo{journal}{\bibinfo{title}{Variance sum rule for entropy
  production}}.
\newblock {\emph{\JournalTitle{Science}}} \textbf{\bibinfo{volume}{383}},
  \bibinfo{pages}{971–976}, \doiprefix\url{10.1126/science.adh1823}
  (\bibinfo{year}{2024}).

\bibitem{Hopfield1974}
\bibinfo{author}{Hopfield, J.~J.}
\newblock \bibinfo{journal}{\bibinfo{title}{Kinetic proofreading: A new
  mechanism for reducing errors in biosynthetic processes requiring high
  specificity}}.
\newblock {\emph{\JournalTitle{Proc. Nat. Acad. Sci.}}}
  \textbf{\bibinfo{volume}{71}}, \bibinfo{pages}{4135–4139},
  \doiprefix\url{10.1073/pnas.71.10.4135} (\bibinfo{year}{1974}).

\bibitem{Andrieux2008}
\bibinfo{author}{Andrieux, D.} \& \bibinfo{author}{Gaspard, P.}
\newblock \bibinfo{journal}{\bibinfo{title}{Nonequilibrium generation of
  information in copolymerization processes}}.
\newblock {\emph{\JournalTitle{Proc. Nat. Acad. Sci.}}}
  \textbf{\bibinfo{volume}{105}}, \bibinfo{pages}{9516–9521},
  \doiprefix\url{10.1073/pnas.0802049105} (\bibinfo{year}{2008}).

\bibitem{Lan2012}
\bibinfo{author}{Lan, G.}, \bibinfo{author}{Sartori, P.},
  \bibinfo{author}{Neumann, S.}, \bibinfo{author}{Sourjik, V.} \&
  \bibinfo{author}{Tu, Y.}
\newblock \bibinfo{journal}{\bibinfo{title}{The energy–speed–accuracy
  trade-off in sensory adaptation}}.
\newblock {\emph{\JournalTitle{Nature Phys.}}} \textbf{\bibinfo{volume}{8}},
  \bibinfo{pages}{422–428}, \doiprefix\url{10.1038/nphys2276}
  (\bibinfo{year}{2012}).

\bibitem{Barato2015}
\bibinfo{author}{Barato, A.~C.} \& \bibinfo{author}{Seifert, U.}
\newblock \bibinfo{journal}{\bibinfo{title}{Thermodynamic uncertainty relation
  for biomolecular processes}}.
\newblock {\emph{\JournalTitle{Phys. Rev. Lett.}}}
  \textbf{\bibinfo{volume}{114}},
  \doiprefix\url{10.1103/physrevlett.114.158101} (\bibinfo{year}{2015}).

\bibitem{dechant2022minimum}
\bibinfo{author}{Dechant, A.}
\newblock \bibinfo{journal}{\bibinfo{title}{Minimum entropy production,
  detailed balance and {Wasserstein} distance for continuous-time markov
  processes}}.
\newblock {\emph{\JournalTitle{Journal of Physics A: Mathematical and
  Theoretical}}} \textbf{\bibinfo{volume}{55}}, \bibinfo{pages}{094001},
  \doiprefix\url{10.1088/1751-8121/ac4ac0} (\bibinfo{year}{2022}).

\bibitem{yoshimura2023housekeeping}
\bibinfo{author}{Yoshimura, K.}, \bibinfo{author}{Kolchinsky, A.},
  \bibinfo{author}{Dechant, A.} \& \bibinfo{author}{Ito, S.}
\newblock \bibinfo{journal}{\bibinfo{title}{Housekeeping and excess entropy
  production for general nonlinear dynamics}}.
\newblock {\emph{\JournalTitle{Physical Review Research}}}
  \textbf{\bibinfo{volume}{5}}, \bibinfo{pages}{013017},
  \doiprefix\url{10.1103/PhysRevResearch.5.013017} (\bibinfo{year}{2023}).

\bibitem{Vu2023}
\bibinfo{author}{Vu, T.~V.} \& \bibinfo{author}{Saito, K.}
\newblock \bibinfo{journal}{\bibinfo{title}{Thermodynamic unification of
  optimal transport: Thermodynamic uncertainty relation, minimum dissipation,
  and thermodynamic speed limits}}.
\newblock {\emph{\JournalTitle{Phys. Rev. X}}} \textbf{\bibinfo{volume}{13}},
  \bibinfo{pages}{011013}, \doiprefix\url{10.1103/physrevx.13.011013}
  (\bibinfo{year}{2023}).

\bibitem{Klinger2025}
\bibinfo{author}{Klinger, J.} \& \bibinfo{author}{Rotskoff, G.~M.}
\newblock \bibinfo{journal}{\bibinfo{title}{Universal energy-speed-accuracy
  trade-offs in driven nonequilibrium systems}}.
\newblock {\emph{\JournalTitle{Phys. Rev. E}}} \textbf{\bibinfo{volume}{111}},
  \doiprefix\url{10.1103/physreve.111.014114} (\bibinfo{year}{2025}).

\bibitem{Sagawa2014}
\bibinfo{author}{Sagawa, T.}
\newblock \bibinfo{journal}{\bibinfo{title}{Thermodynamic and logical
  reversibilities revisited}}.
\newblock {\emph{\JournalTitle{Journal of Statistical Mechanics: Theory and
  Experiment}}} \textbf{\bibinfo{volume}{2014}}, \bibinfo{pages}{P03025},
  \doiprefix\url{10.1088/1742-5468/2014/03/p03025} (\bibinfo{year}{2014}).

\bibitem{Schmiedl2007}
\bibinfo{author}{Schmiedl, T.} \& \bibinfo{author}{Seifert, U.}
\newblock \bibinfo{journal}{\bibinfo{title}{Optimal finite-time processes in
  stochastic thermodynamics}}.
\newblock {\emph{\JournalTitle{Phys. Rev. Lett.}}}
  \textbf{\bibinfo{volume}{98}}, \bibinfo{pages}{108301},
  \doiprefix\url{10.1103/physrevlett.98.108301} (\bibinfo{year}{2007}).

\bibitem{Blaber2023}
\bibinfo{author}{Blaber, S.} \& \bibinfo{author}{Sivak, D.~A.}
\newblock \bibinfo{journal}{\bibinfo{title}{Optimal control in stochastic
  thermodynamics}}.
\newblock {\emph{\JournalTitle{J. Phys. Comm.}}} \textbf{\bibinfo{volume}{7}},
  \bibinfo{pages}{033001}, \doiprefix\url{10.1088/2399-6528/acbf04}
  (\bibinfo{year}{2023}).

\bibitem{Loos2024}
\bibinfo{author}{Loos, S.~A.}, \bibinfo{author}{Monter, S.},
  \bibinfo{author}{Ginot, F.} \& \bibinfo{author}{Bechinger, C.}
\newblock \bibinfo{journal}{\bibinfo{title}{Universal symmetry of optimal
  control at the microscale}}.
\newblock {\emph{\JournalTitle{Phys. Rev. X}}} \textbf{\bibinfo{volume}{14}},
  \bibinfo{pages}{021032}, \doiprefix\url{10.1103/physrevx.14.021032}
  (\bibinfo{year}{2024}).

\bibitem{Horowitz2019}
\bibinfo{author}{Horowitz, J.~M.} \& \bibinfo{author}{Gingrich, T.~R.}
\newblock \bibinfo{journal}{\bibinfo{title}{Thermodynamic uncertainty relations
  constrain non-equilibrium fluctuations}}.
\newblock {\emph{\JournalTitle{Nature Phys.}}} \textbf{\bibinfo{volume}{16}},
  \bibinfo{pages}{15–20}, \doiprefix\url{10.1038/s41567-019-0702-6}
  (\bibinfo{year}{2019}).

\bibitem{Song2021}
\bibinfo{author}{Song, Y.} \& \bibinfo{author}{Hyeon, C.}
\newblock \bibinfo{journal}{\bibinfo{title}{Thermodynamic uncertainty relation
  to assess biological processes}}.
\newblock {\emph{\JournalTitle{J. Chem. Phys.}}}
  \textbf{\bibinfo{volume}{154}}, \bibinfo{pages}{130901},
  \doiprefix\url{10.1063/5.0043671} (\bibinfo{year}{2021}).

\bibitem{Marsland2019}
\bibinfo{author}{Marsland, R.}, \bibinfo{author}{Cui, W.} \&
  \bibinfo{author}{Horowitz, J.~M.}
\newblock \bibinfo{journal}{\bibinfo{title}{The thermodynamic uncertainty
  relation in biochemical oscillations}}.
\newblock {\emph{\JournalTitle{J. R. Soc. Interface}}}
  \textbf{\bibinfo{volume}{16}}, \bibinfo{pages}{20190098},
  \doiprefix\url{10.1098/rsif.2019.0098} (\bibinfo{year}{2019}).

\bibitem{Ball2012}
\bibinfo{author}{Ball, P.}
\newblock \bibinfo{journal}{\bibinfo{title}{Computer engineering: Feeling the
  heat}}.
\newblock {\emph{\JournalTitle{Nature}}} \textbf{\bibinfo{volume}{492}},
  \bibinfo{pages}{174–176}, \doiprefix\url{10.1038/492174a}
  (\bibinfo{year}{2012}).

\bibitem{Markov2014}
\bibinfo{author}{Markov, I.~L.}
\newblock \bibinfo{journal}{\bibinfo{title}{Limits on fundamental limits to
  computation}}.
\newblock {\emph{\JournalTitle{Nature}}} \textbf{\bibinfo{volume}{512}},
  \bibinfo{pages}{147–154}, \doiprefix\url{10.1038/nature13570}
  (\bibinfo{year}{2014}).

\bibitem{Freitas2021}
\bibinfo{author}{Freitas, N.}, \bibinfo{author}{Delvenne, J.-C.} \&
  \bibinfo{author}{Esposito, M.}
\newblock \bibinfo{journal}{\bibinfo{title}{Stochastic thermodynamics of
  nonlinear electronic circuits: A realistic framework for computing around
  {kT}}}.
\newblock {\emph{\JournalTitle{Phys. Rev. X}}} \textbf{\bibinfo{volume}{11}},
  \doiprefix\url{10.1103/physrevx.11.031064} (\bibinfo{year}{2021}).

\bibitem{Wolpert2024}
\bibinfo{author}{Wolpert, D.~H.} \emph{et~al.}
\newblock \bibinfo{journal}{\bibinfo{title}{Is stochastic thermodynamics the
  key to understanding the energy costs of computation?}}
\newblock {\emph{\JournalTitle{Proc. Nat. Acad. Sci.}}}
  \textbf{\bibinfo{volume}{121}}, \doiprefix\url{10.1073/pnas.2321112121}
  (\bibinfo{year}{2024}).

\end{thebibliography}

\begin{thebibliography}{1}
\urlstyle{rm}
\expandafter\ifx\csname url\endcsname\relax
  \def\url#1{\texttt{#1}}\fi
\expandafter\ifx\csname urlprefix\endcsname\relax\def\urlprefix{URL }\fi
\expandafter\ifx\csname doiprefix\endcsname\relax\def\doiprefix{DOI: }\fi
\providecommand{\bibinfo}[2]{#2}
\providecommand{\eprint}[2][]{\url{#2}}

\bibitem{Villani2009}
\bibinfo{author}{Villani, C.}
\newblock \emph{\bibinfo{title}{Optimal transport: old and new}}
  (\bibinfo{publisher}{Springer}, \bibinfo{year}{2009}).

\bibitem{flamary2021pot}
\bibinfo{author}{Flamary, R.} \emph{et~al.}
\newblock \bibinfo{journal}{\bibinfo{title}{{POT: Python Optimal Transport}}}.
\newblock {\emph{\JournalTitle{J. Mach. Learn. Res.}}}
  \textbf{\bibinfo{volume}{22}}, \bibinfo{pages}{1--8} (\bibinfo{year}{2021}).

\bibitem{Benamou2000}
\bibinfo{author}{Benamou, J.-D.} \& \bibinfo{author}{Brenier, Y.}
\newblock \bibinfo{journal}{\bibinfo{title}{A computational fluid mechanics
  solution to the {Monge--Kantorovich} mass transfer problem}}.
\newblock {\emph{\JournalTitle{Num. Math.}}} \textbf{\bibinfo{volume}{84}},
  \bibinfo{pages}{375–393}, \doiprefix\url{10.1007/s002110050002}
  (\bibinfo{year}{2000}).

\bibitem{Seifert2012}
\bibinfo{author}{Seifert, U.}
\newblock \bibinfo{journal}{\bibinfo{title}{Stochastic thermodynamics,
  fluctuation theorems and molecular machines}}.
\newblock {\emph{\JournalTitle{Rep. Prog. Phys.}}}
  \textbf{\bibinfo{volume}{75}}, \bibinfo{pages}{126001},
  \doiprefix\url{10.1088/0034-4885/75/12/126001} (\bibinfo{year}{2012}).

\bibitem{Pigolotti-Peliti}
\bibinfo{author}{Peliti, L.} \& \bibinfo{author}{Pigolotti, S.}
\newblock \emph{\bibinfo{title}{Stochastic Thermodynamics: An Introduction}}
  (\bibinfo{publisher}{Princeton University Press}, \bibinfo{year}{2021}).

\bibitem{Nakazato2021}
\bibinfo{author}{Nakazato, M.} \& \bibinfo{author}{Ito, S.}
\newblock \bibinfo{journal}{\bibinfo{title}{Geometrical aspects of entropy
  production in stochastic thermodynamics based on {Wasserstein} distance}}.
\newblock {\emph{\JournalTitle{Phys. Rev. Res.}}} \textbf{\bibinfo{volume}{3}},
  \bibinfo{pages}{043093}, \doiprefix\url{10.1103/physrevresearch.3.043093}
  (\bibinfo{year}{2021}).

\bibitem{Schmiedl2007}
\bibinfo{author}{Schmiedl, T.} \& \bibinfo{author}{Seifert, U.}
\newblock \bibinfo{journal}{\bibinfo{title}{Optimal finite-time processes in
  stochastic thermodynamics}}.
\newblock {\emph{\JournalTitle{Phys. Rev. Lett.}}}
  \textbf{\bibinfo{volume}{98}}, \bibinfo{pages}{108301},
  \doiprefix\url{10.1103/physrevlett.98.108301} (\bibinfo{year}{2007}).

\end{thebibliography}

\begin{thebibliography}{1}
\urlstyle{rm}
\expandafter\ifx\csname url\endcsname\relax
  \def\url#1{\texttt{#1}}\fi
\expandafter\ifx\csname urlprefix\endcsname\relax\def\urlprefix{URL }\fi
\expandafter\ifx\csname doiprefix\endcsname\relax\def\doiprefix{DOI: }\fi
\providecommand{\bibinfo}[2]{#2}
\providecommand{\eprint}[2][]{\url{#2}}

\bibitem{2020SciPy-NMeth}
\bibinfo{author}{Virtanen, P.} \emph{et~al.}
\newblock \bibinfo{journal}{\bibinfo{title}{{{SciPy} 1.0: Fundamental
  Algorithms for Scientific Computing in Python}}}.
\newblock {\emph{\JournalTitle{Nature Methods}}} \textbf{\bibinfo{volume}{17}},
  \bibinfo{pages}{261--272}, \doiprefix\url{10.1038/s41592-019-0686-2}
  (\bibinfo{year}{2020}).

\bibitem{Villani2009}
\bibinfo{author}{Villani, C.}
\newblock \emph{\bibinfo{title}{Optimal transport: old and new}}
  (\bibinfo{publisher}{Springer}, \bibinfo{year}{2009}).

\bibitem{Klinger2025}
\bibinfo{author}{Klinger, J.} \& \bibinfo{author}{Rotskoff, G.~M.}
\newblock \bibinfo{journal}{\bibinfo{title}{Universal energy-speed-accuracy
  trade-offs in driven nonequilibrium systems}}.
\newblock {\emph{\JournalTitle{Phys. Rev. E}}} \textbf{\bibinfo{volume}{111}},
  \doiprefix\url{10.1103/physreve.111.014114} (\bibinfo{year}{2025}).

\bibitem{Proesmans2020}
\bibinfo{author}{Proesmans, K.}, \bibinfo{author}{Ehrich, J.} \&
  \bibinfo{author}{Bechhoefer, J.}
\newblock \bibinfo{journal}{\bibinfo{title}{Finite-time {Landauer} principle}}.
\newblock {\emph{\JournalTitle{Phys. Rev. Lett.}}}
  \textbf{\bibinfo{volume}{125}}, \bibinfo{pages}{100602},
  \doiprefix\url{10.1103/physrevlett.125.100602} (\bibinfo{year}{2020}).

\bibitem{Proesmans_2020-2}
\bibinfo{author}{Proesmans, K.}, \bibinfo{author}{Ehrich, J.} \&
  \bibinfo{author}{Bechhoefer, J.}
\newblock \bibinfo{journal}{\bibinfo{title}{Optimal finite-time bit erasure
  under full control}}.
\newblock {\emph{\JournalTitle{Phys. Rev. E}}} \textbf{\bibinfo{volume}{102}},
  \bibinfo{pages}{032105}, \doiprefix\url{10.1103/physreve.102.032105}
  (\bibinfo{year}{2020}).

\end{thebibliography}
\end{document}